\documentclass[journal,10pt,twoside]{IEEEtran}

\usepackage[
  letterpaper,
  left=0.680in,
  right=0.680in,
  top=0.811in,
  bottom=0.557in,
  headheight=14pt,
  headsep=14pt,
  columnsep=0.167in
]{geometry}

\usepackage{amsmath,amsfonts,cases,amsthm,cite,url,graphicx,tikz,subfigure,hyperref}

\usepackage{xcolor,etoolbox}

\newtheorem{definition}{Definition}
\newtheorem{corollary}{Corollary}
\newtheorem{theorem}{Theorem}
\newtheorem{remark}{Remark}
\newtheorem{lemma}{Lemma}

\usetikzlibrary{graphs}

\DeclareMathOperator{\EX}{\mathbb{E}}

\makeatletter

\def\env@sqcases{%
  \let\@ifnextchar\new@ifnextchar
  \left\lbrack
  \def\arraystretch{1.2}%
  \array{@{}l@{\quad}l@{}}%
}
\makeatother

\newcommand{\T}{\mathcal{T}_{\epsilon}}
\newcommand{\Err}{\mathcal{E}}

\newcommand{\set}{\newcommand}

\set{\RS}{R_{1}}
\set{\RSc}{ {R_{\color{blue}i}} }
\set{\RSp}{ {R_{\color{blue}j}} }
\set{\sck}{a_0^k(i)}
\set{\spk}{a_1^k(i,j)}
\set{\Sc}{A_0} \set{\Sp}{A_1}
\set{\A}{A}

\set{\RU}{R_{2}}
\set{\RUc}{ {R_{\color{blue}l}} }
\set{\RUp}{ {R_{\color{blue}p}} }

\set{\uck}{b_0^k(i,l)}
\set{\upk}{b_1^k(i,l,p)}
\set{\Uc}{B_0} \set{\Up}{B_1}
\set{\B}{B}

\set{\ws}{m_1'} \set{\wu}{m_2'}

\set{\wsd}{\hat{m}_1'} \set{\wud}{\hat{m}_2'}

\set{\Acc}{Q_1} \set{\Bcc}{Q_2}
\set{\Bb}{B'}
\set{\App}{W_1} \set{\Bpp}{W_2}
\set{\Aa}{A'}
\set{\ABcc}{Q} \set{\ABpp}{W}

\set{\RAcc}{ {R'_{\color{blue}i}} } \set{\RBcc}{ R'_{\color{blue}l} }
\set{\RApp}{ {R'_{\color{blue}j}} } \set{\RBpp}{ {R'_{\color{blue}p}} }
\set{\Rms}{R'_{m1}}
\set{\Rmu}{R'_{m2}}

\set{\qsns}{{\color{blue} q_1^n}}
\set{\quns}{{\color{blue} q_2^n}}
\set{\wsns}{{\color{blue} w_1^n}}
\set{\wuns}{{\color{blue} w_2^n}}

\set{\qsn}{\qsns {\color{blue}(i')}}
\set{\wsn}{\wsns {\color{blue}(i',j',\ws)}}
\set{\qun}{\quns {\color{blue}(i',l')}}
\set{\wun}{\wuns {\color{blue}(i',l',p',\wu)}}

\set{\qsnd}{\qsns(\hat{i}')}
\set{\wsnd}{\wsns(\hat{i}',\hat{j}',\wsd)}
\set{\qund}{\quns(\hat{i}',\hat{l}')}
\set{\wund}{\wuns(\hat{i}',\hat{l}',\hat{p}',\wud)}

\set{\Ss}{V} \set{\U}{Q} \set{\V}{W}

\begin{document}

\let\OldColor\color
\renewcommand{\color}[1]{\OldColor{black}}

\title{
Secure Semantic Communication Over Wiretap Channels: Rate-Distortion-Equivocation Tradeoff

\thanks{
This work was supported by the 5GIC
and 6GIC, Institute for Communication Systems (ICS), University of Surrey.
An earlier version of this paper was presented in part at the 2024 IEEE Information Theory Workshop (ITW) [DOI: 10.1109/ITW61385.2024.10807039].
\textit{(Corresponding author: Denis Kozlov.) \nocite{kozlov2024}}

The authors are with the 5GIC and 6GIC, Institute for Communication Systems (ICS), University of Surrey, GU2 7XH Guildford, U.K. (e-mail: d.kozlov@surrey.ac.uk; m.mirmohseni@surrey.ac.uk;
r.tafazolli@surrey.ac.uk).
}}

\author{\IEEEauthorblockN{
Denis Kozlov, \textit{Student Member, IEEE},
Mahtab Mirmohseni, \textit{Senior Member, IEEE}, \\
and Rahim Tafazolli \textit{Fellow, IEEE}
}}

\markboth{Kozlov \MakeLowercase{\textit{et al.}}: Secure Semantic Communication Over Wiretap Channels}{}

 
\maketitle

\ifCLASSOPTIONdraft
\vspace{-50pt}
\fi
\begin{abstract} 
This paper investigates an information-theoretic model of secure semantic-aware communication. For this purpose, we consider the lossy joint source-channel coding (JSCC) of a memoryless semantic source transmitted over a memoryless wiretap channel. The source consists of two correlated parts that represent semantic and observed aspects of the information. Our model assumes separate fidelity and secrecy constraints on each source component and, in addition, encompasses two cases for the source output, in order to evaluate the performance gains if the encoder has an extended access to the source. Specifically, in Case 1, the encoder has direct access only to the samples from a single (observed) source component, while in Case 2 it has additional direct access to the samples of the underlying semantic information. We derive single-letter converse and achievability bounds on the rate-distortion-equivocation region. The converse bound explicitly contains rate-distortion functions, making it easy to evaluate, especially for some common distributions. The proposed achievability coding scheme involves novel stochastic superposition coding with two private parts to enable analysis of the equivocation for each source component, separately. Our results generalise some of the previously established source and source-channel coding problems. The general results are further specialised to Gaussian and Bernoulli sources transmitted over Gaussian and binary wiretap channels, respectively. The numerical evaluations illustrate the derived bounds for these distributions.
\end{abstract}

\section{Introduction}
\IEEEPARstart{T}{raditional} communication systems are {\color{blue} increasingly constrained by classical information-theoretic design paradigms,} in part due to a growing volume of information exchange from diverse sources, driven by a widespread integration of machine learning (ML) in numerous applications. This motivates us to use new options like semantic communication, which is a promising approach for next-generation wireless networks, especially within the realm of 6G technology \cite{aazhang2019, yang2023}.

The semantic and accompanying concept of task-orientated communication can be implemented using conventional communication systems, in part by modifying the underlying compression algorithm. However, the joint source-channel coding (JSCC) offers many advantages over separate source-channel coding systems, making it a particularly suitable foundation for semantic communication systems \cite{gunduz2024jscc}. While recent advances in JSCC have addressed semantic-aware transmission, both from the applied \cite{bourtsoulatze2019, tung2021, gunduz2024jscc} and theoretical \cite{stavrou2023, stavrou2023jscc, gunduz2024jscc} perspectives, \textit{secure} communication of semantic data remains understudied, particularly from an information-theoretic perspective. Preserving security in semantic communication has unique challenges because semantic communication still faces substantial security challenges due to the inherent openness of communication channels, leaving the semantics of the data susceptible to eavesdropping. Furthermore, semantics can carry more sensitive information compared to conventional data. Thus, security requirements can be different for them. In addition, in many applications, it is necessary to protect only specific semantic features, rather than entire data, and securing all data may be inefficient and unnecessary. Therefore, it is important to explore the fundamental limits that govern such selective protection of semantic information.

To this end, we investigate secure semantic-aware communication from an information-theoretic point of view, in order to characterise its fundamental limits. For this purpose, we consider a model of a secure lossy JSCC of semantic sources, transmitted over a wiretap channel. In contrast to a classic information-theoretic model of an information source, which is modelled by a single component, i.e., by a random variable (r.v.)\footnote{In a memoryless case.}, the semantic source \cite{liu2021indirect, liu2022indirect} is modelled by at least two r.v.s, $S$ and $U$, representing the semantic and observed parts of the data, respectively. This semantic source model is studied in various information-theoretic settings \cite{liu2022indirect, guo2022, guo2022rdf, xiao2022, stavrou2023jscc, chai2023}. However, \textit{secure} JSCC of semantic sources remains understudied. In such case, the objective of the encoder is not only to guarantee fidelity for both the semantic and observed parts of the semantic source at the legitimate receiver, but also to maintain separate secrecy constraints on each source component at the eavesdropper's side.  These distinct secrecy constraints make the model adaptable to a variety of semantic communication scenarios. A typical case is one where maximum secrecy is required for semantic information, while the secrecy of observed data is considered irrelevant. The problem is to characterise the trade-off between communication rate, fidelity, and secrecy within the JSCC model. Wherein, we assume that secrecy has three degrees of freedom, which should be balanced based on available resources: i) lossy compression, which limits fidelity of the data, ii) encryption with limited shared secret key resource, and iii) channel secrecy capacity rate, limited by the wiretap channel statistics.

\subsection{Related Works}
The classic source model, where the message to be sent is modelled as a single r.v., is studied in a variety of secure communication settings, with  Wyner's work \cite{wyner1975} in 1975 laying the groundwork for secure communication over wiretap channels. The core idea described in \cite{wyner1975} is that the difference in channel statistics between the legitimate receiver and the eavesdropper can be exploited in terms of secrecy with the help of the appropriate stochastic coding without the use of encryption. The subsequent advancements generalised this model to broadcast channels with common and confidential messages \cite{csiszar1978}.
The authors of \cite{wyner1975,csiszar1978} exclusively study \textit{channel coding} problem. Whereas, later in \cite{yamamoto1997}, the wiretap model is extended to incorporate lossy JSCC and the one-time pad (OTP) encryption technique, bringing an important result that the separation principle holds for secure JSCC: one can start with a rate-distortion achievable code, followed by a encryption for a given key rate, and finish by using a wiretap code. The author of \cite{yamamoto1997} shows that within his model secrecy constitutes the trade-off between rate, distortion, secret key rate, and secrecy capacity.
Further extensions of the secure JSCC model, including scenarios with source-correlated side-information at the decoders, are explored in \cite{merhav2007, villard2014, villard2011phd}. In \cite{merhav2007} it is shown that if the eavesdropper has degraded channel and side-information with respect to the legitimate receiver, the separation of the source and channel coding is asymptotically optimal. The optimal approach in this case is to use Wyner-Ziv coding \cite{wyner-ziv1976} followed by encryption and wiretap channel coding. In \cite{villard2014}, the model of \cite{merhav2007} is generalised to the case of a non-necessary degraded channel and side-information. In general, this relaxation results in non-optimality of source-channel coding separation except for some specific cases. However, the authors of \cite{villard2014} show the important result -- secrecy can be achieved with the help of side-information, even if the eavesdropper has better channel statistics than the legitimate receiver.

The foundational model of semantic source and its rate-distortion description are introduced in \cite{liu2022indirect}. Compared to the classic source model, the semantic source model includes an additional r.v. to encompass semantic aspects of the information. 
Several works have investigated this model in different settings. For example, in \cite{guo2022rdf}, the semantic source coding problem is extended to include side-information at the decoder.
In parallel, in \cite{stavrou2023jscc}, the authors considered the transmission of a semantic source over the noisy channel and derived sufficient conditions for the optimal transmission in the case of lossy JSCC. Their work can be seen as an adaptation of the celebrated results \cite{gastpar2000} to the semantic source model.
The authors of \cite{chai2023} extend the source coding problem to the one with perception constraint and study the rate-distortion-perception function for the semantic source. The semantic source model is also considered in game-theoretic settings \cite{xiao2022}, and in \cite{li2023}, it is placed in a practical scenario. Specifically, the authors of \cite{li2023} cover the case of unknown source distribution and propose ML-based algorithm to estimate semantic source distribution.

To the best of our knowledge, the first work that considers a source model similar to the semantic source in a \textit{secure} setting is \cite{yamamoto1994}. Specifically, the author of \cite{yamamoto1994} studies the problem of \textit{lossless secure source coding} of two correlated source outputs within Shannon’s cipher system. The paper lists possible transmission cases covering whether one part or both should be secret or whether one part or both are transmitted. For each case, they show an admissible region for the data rate and secret key rate.
Previously, in \cite{yamamoto1983}, the same author studied the model of \textit{secure lossy source coding} in which the distortion constraint is imposed on one source component (in our model it corresponds to the observed part) while the secrecy constraint is imposed on the other (semantic part in our model). In \cite{yamamoto1983}, the model does not assume a secret key and secrecy is achieved only by compression. The author presents a rate-distortion-equivocation tradeoff for two cases of encoder input (in a similar way to our definition of Case 1 and Case 2 encoding, see Section~\ref{sec:problem}).
The \textit{secure lossy source coding with secret key} of semantic sources is studied in \cite{guo2022}, where the secrecy condition is imposed only on the semantic part of the data while the fidelity (distortion) constraints are considered for each source component. The work \cite{guo2022} presents a tight bound for the rate-distortion-equivocation region.
The interested reader may also find results for relevant secure source coding problems in \cite{benammar2016, benammar2017}, including variations of Gray-Wyner problems with secrecy constraints.

\subsection{Contributions and Paper Structure}
The contributions of this paper can be summarised in the following.
    1) We present a novel information-theoretic model for secure semantic-aware communication. The model follows the lossy JSCC principle and assumes separate fidelity (distortion) and secrecy constraints on the two correlated source components. In our model, we consider optional encryption with limited secret-key resource, that allows to study fundamental limits of semantic-aware communication in combination with encryption techniques. The model also assumes two scenarios for semantic source encoding, which allow us to compare performance gains if the encoder has extended view on the source.
    2) For our model, we characterise trade-offs between communication rate, fidelity, and secrecy. Specifically, we derive a single-letter converse and achievability bounds on the rate-distortion-equivocation region. Our converse region explicitly contains rate-distortion functions (RDFs) and secrecy capacity terms, which simplifies its numerical evaluation as there exist closed-from solutions for some common source and channel distributions. The proposed achievability scheme involves novel approach allowing to separately have control over equivocation in addition to two separate distortions for two correlated parts of the source. 
    3) We compare our model and findings with some known formulations and show that our work generalises some of them. Specifically, our model can be reduced to lossy secure JSCC of \textit{classic sources} \cite{yamamoto1997} and to secure \textit{source coding} of semantic sources \cite{yamamoto1983, guo2022}.
    4) We reduce our general converse and achievability bound to the Gaussian and Bernoulli sources and Gaussian and binary channels to numerically present our region. The comparison for different encoder setups is presented and performance gain is highlighted.

In this article, we study the same model that we presented at the conference \cite{kozlov2024}. Compared to \cite{kozlov2024}, in this work, we present extended results, including {\color{blue} the generalization to not necessarily degraded wiretap channel,} tighter secrecy capacity terms in the converse, tighter achievability bound, which supports {\color{blue} $R_k \geq 0$,} Case 1 and Case 2 of the encoder output (see problem statement in Section~\ref{sec:problem} for definitions of Case 1 and Case 2) and improved numerical results which also append the binary case of the source and channel. We compare our results with some of the previous works and show that our results generalises some of them. We also provide a detailed discussion and comparison of the results.

The paper is organised as follows. In Section~\ref{sec:problem}, we rigorously describe the problem at hand and present some supplementary results from the rate-distortion theory. Section~\ref{sec:converse-result} and Section~\ref{sec:ach-result} present the main results of this paper, describing the fundamental limits of our model: converse and achievable bounds on the rate-distortion-equivocation region. Section~\ref{sec:gauss-and-binary} is dedicated to reduced results for the cases of the Gaussian and Bernoulli (binary) sources and channels. This section also contains a numerical evaluation of rate-distortion-equivocation trade-offs. Section~\ref{sec:conclusion} concludes this work. This paper also {\color{teal}has supplementary material consisting of }Appendix~\ref{appendix:cardinality}{\color{blue}-\ref{appendix:mi-term}} to support some elements of the {\color{blue} converse and} achievability proofs.
\subsection*{Notation}
All logarithm functions within this paper are in base 2. The standard notation $\log^+(x)$ means $\max \left[ 0, log(x) \right]$ and $[x]^+$ means $\max \left[ 0, x \right]$. $\EX(X)$ is the expected value of $X$. {\color{blue} We denote binary XOR operation with $\oplus$}. The discrete entropy function is denoted as $H(.)$, the binary entropy function shown as $H_b(.)$, and the differential entropy as $h(.)$. We define: $a * b \doteq a(1 - b) + (1 - a)b$. We denote random variable (r.v.) with capital letters (e.g. $X$), the lowercase equivalent represents the realisation of r.v. (e.g. $x$) and the capital letter in calligraphic font stands for the alphabet of r.v. (e.g. $\mathcal{X}$). In our work, we refer to $S$ as the semantic part and $U$ as the observed part of the semantic source, the reconstruction r.v.s for the semantic and observed part denoted as $\hat{S}$ and $\hat{U}$, respectively. $X,Y,Z$ represent channel r.v.s. The probability mass function of some r.v. $X$ is written as $p_X(x)$ or for brevity as $p_X$. The superscript $k$ denotes the source sequence of length $k$, while $n$ stands for the channel sequence of length $n$. Strongly typical and conditionally typical sets of $n$-lengthened sequences are denoted $\T^n(X)$ and $\T^n(X|y)$, respectively. We write a normal distribution using $\mathcal{N}(.,.)$ and a Bernoulli distribution using $\mathcal{B}(.)$. We denote the indicator function of $x$ and $\hat{x}$ as $\mathbb{I}_{\{x \neq \hat{x}\}}$. The determinant of matrix $\mathbf{M}$ is denoted as $|\mathbf{M}|$. We refer to equation number when it is used as a superscript in our derivations, e.g. $=^{(42)}$ means that the equality is based on equation (42).
\section{Problem Statement and Supplementary Results}
\label{sec:problem}
\subsection{Problem Statement}

We consider the communication model shown in Fig.~\ref{fig:model}, where a JSCC transmitter aims to communicate $k$ independent and identically distributed (i.i.d.) samples of both semantic, $S^k$, and observed, $U^k$, data to a legitimate receiver under two key requirements: i) maintaining separate average distortion constraints for each information component at the legitimate receiver, and ii) ensuring information-theoretic secrecy against an eavesdropper through separate equivocation constraints. The correlated source components $S^k$ (semantic) and $U^k$ (observed) are modelled as sequences of i.i.d. r.v.s with joint distribution $p_{S,U}$ over the product alphabet $\mathcal{S} \times \mathcal{U}$.

\ifCLASSOPTIONdraft
\begin{figure}
\centering
\includegraphics[draft=False,width=1.0\linewidth]{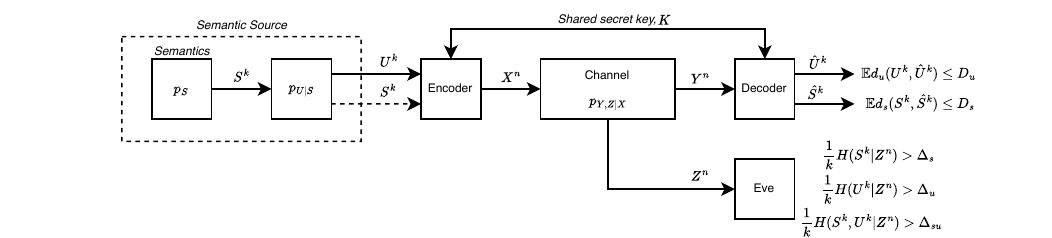}
\caption{Wiretap channel model for semantic communication with two encoder configurations: Case 1 (observation-only input) and Case 2 (joint semantic-observation input). System objectives include dual distortion constraints at the legitimate receiver ($D_s, D_u$) and triple equivocation constraints at the eavesdropper ($\Delta_s, \Delta_u, \Delta_{su}$).}
\label{fig:model}
\end{figure}
\else
\begin{figure}
\centering
\includegraphics[width=1.0\linewidth]{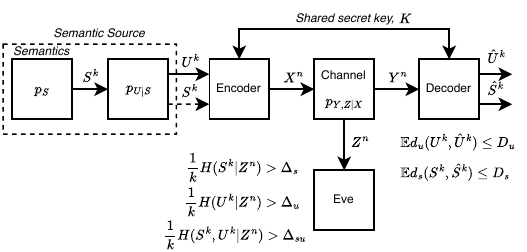}
\caption{Wiretap channel model for semantic communication with two encoder configurations: Case 1 (observation-only input) and Case 2 (joint semantic-observation input). System objectives include dual distortion constraints at the legitimate receiver ($D_s, D_u$) and triple equivocation constraints at the eavesdropper ($\Delta_s, \Delta_u, \Delta_{su}$).}
\label{fig:model}
\end{figure}
\fi

The main and the wiretap channels are modelled as discrete memoryless channels (DMCs) governed by the transition probability $p_{Y,Z|X}$, with input alphabet $\mathcal{X}$ and output alphabets $\mathcal{Y}$ and $\mathcal{Z}$. Our model also assumes a rate-constrained shared secret key $K \sim p_K$ shared between the encoder and the legitimate receiver, defined on the alphabet $\mathcal{K}$ independent of the source and channel r.v.s to enable partial secrecy through encryption.

To assess the performance gain we get if the encoder is directly aware of the semantic part samples, we consider the following two scenarios:

\begin{definition}
[Case 1: observation-only encoding]
Having access only to sequence $U^k$ and secret key $K$, the encoder $f_1: \mathcal{U}^k \times \mathcal{K} \rightarrow \mathcal{X}^n$ encodes $U^k$ into the channel input sequence $X^n$.
\end{definition}

\begin{definition}
[Case 2: joint semantic-observation encoding]
The encoder $f_2: \mathcal{S}^k \times \mathcal{U}^k \times \mathcal{K} \rightarrow \mathcal{X}^n$ given the secret key $K$ jointly encodes both semantic and observed sequences $S^k$, $U^k$ into $X^n$.
\end{definition}

\begin{definition}[Decoder] \label{def:decoder}
The decoder $\hat{f} = (\hat{f}_s, \hat{f}_u)$, where $\hat{f}_s: \mathcal{Y}^n \times \mathcal{K} \rightarrow \hat{\mathcal{S}}^k$ and $\hat{f}_u: \mathcal{Y}^n \times \mathcal{K} \rightarrow \hat{\mathcal{U}}^k$, reconstructs estimates $(\hat{S}^k, \hat{U}^k)$ using the channel output $Y^n$ and the secret key $K$. 
\end{definition}

The single-letter distortion is quantified through the functions $d_s: \mathcal{S} \times \hat{\mathcal{S}} \rightarrow [0, \infty)$ and $d_u: \mathcal{U} \times \hat{\mathcal{U}} \rightarrow [0, \infty)$ for the semantic and observed parts, respectively. The block distortion averages are denoted as follows:\ifCLASSOPTIONdraft
\begin{align}
    d_s(s^k,\hat{s}^k) \doteq \frac{1}{k}\sum_{i=1}^k d_s(s_i,\hat{s}_i), \quad
    d_u(u^k,\hat{u}^k) \doteq \frac{1}{k}\sum_{i=1}^k d_u(u_i,\hat{u}_i).
\end{align}
\else
\vspace{-0.1cm}
\begin{align}
    &d_s(s^k,\hat{s}^k) \doteq \frac{1}{k}\sum_{i=1}^k d_s(s_i,\hat{s}_i),
\end{align}
\begin{align}
    &d_u(u^k,\hat{u}^k) \doteq \frac{1}{k}\sum_{i=1}^k d_u(u_i,\hat{u}_i).
\end{align}
\fi

\begin{definition}[Code] \label{def:code}
A $(k,n)$ source-channel code is defined by a stochastic encoding function $f_i$ and a decoding function $\hat{f}$, where $i \in \{1,2\}$ denotes the encoder input case.
\end{definition}

Our objective is to provide bounds for the achievable region \( \mathcal{R} \) of \( (r, R_k, D_s, D_u, \Delta_s, \Delta_u, \Delta_{su}) \), where:
\begin{itemize}
    \item \( r \): channel uses per source symbol,
    \item \( R_k \): secret key rate shared between encoder and the legitimate decoder,
    \item \( D_s, D_u \): semantic/observation distortion constraints,
    \item \( \Delta_s, \Delta_u, \Delta_{su} \): semantic, observation, and joint equivocation thresholds.
\end{itemize}

\begin{definition}[Achievable Tuple of Region Parameters] \label{def:achievable}
A tuple \( (r, R_k, D_s, D_u, \Delta_s, \Delta_u, \Delta_{su}) \in \mathbb{R}_+^7 \) is achievable if for any \( \epsilon > 0 \), there exists a \( (k,n) \)-code satisfying:
\ifCLASSOPTIONdraft
\begin{align}
    \frac{n}{k} &\leq r + \epsilon
    \hspace{7mm} {\text{\rm(rate constraint)}}
    \label{ineq:rate-cond} \\
    \frac{1}{k}\log|\mathcal{K}| &\leq R_k + \epsilon
    \hspace{5mm} {\text{\rm(key rate constraint)}}
    \label{eq:key-rate} \\
    \EX d_s(S^k, \hat{S}^k) &\leq D_s + \epsilon
    \hspace{5mm} {\text{\rm(semantic part distortion)}} 
    \label{eq:sem-distortion} \\
    \EX d_u(U^k, \hat{U}^k) &\leq D_u + \epsilon
    \hspace{5mm} {\text{\rm(observed part distortion)}} 
    \label{eq:obs-distortion} \\
    \frac{1}{k}H(S^k|Z^n) &\geq \Delta_s - \epsilon
    \hspace{5mm} {\text{\rm(semantic part secrecy)}}
    \label{ineq:sem-eqv} \\
    \frac{1}{k}H(U^k|Z^n) &\geq \Delta_u - \epsilon
    \hspace{5mm} {\text{\rm(observed part secrecy)}}
    \label{ineq:obs-eqv} \\
    \frac{1}{k}H(S^k,U^k|Z^n) &\geq \Delta_{su} - \epsilon
    \hspace{4mm} {\text{\rm(joint secrecy)}}
    \label{ineq:joint-eqv}
\end{align}
\else
\vspace{-0.2cm}
\begin{align}
    \frac{n}{k} &\leq r + \epsilon
    \hspace{5mm} {\text{\rm(rate constraint)}}
    \label{ineq:rate-cond} \\
    \frac{1}{k}\log|\mathcal{K}| &\leq R_k + \epsilon
    \hspace{3mm} {\text{\rm(key rate constraint)}}
    \label{eq:key-rate} \\
    \EX d_s(S^k, \hat{S}^k) &\leq D_s + \epsilon
    \hspace{3mm} {\text{\rm(sem. part distortion)}} 
    \label{eq:sem-distortion} \\
    \EX d_u(U^k, \hat{U}^k) &\leq D_u + \epsilon
    \hspace{3mm} {\text{\rm(obs. part distortion)}} 
    \label{eq:obs-distortion} \\
    \frac{1}{k}H(S^k|Z^n) &\geq \Delta_s - \epsilon
    \hspace{3mm} {\text{\rm(sem. part secrecy)}}
    \label{ineq:sem-eqv} \\
    \frac{1}{k}H(U^k|Z^n) &\geq \Delta_u - \epsilon
    \hspace{3mm} {\text{\rm(obs. part secrecy)}}
    \label{ineq:obs-eqv} \\
    \frac{1}{k}H(S^k,U^k|Z^n) &\geq \Delta_{su} - \epsilon
    \hspace{2mm} {\text{\rm(joint secrecy)}}
    \label{ineq:joint-eqv}
\end{align}
\fi
\end{definition}

The first inequality (\ref{ineq:rate-cond}) restricts the number of channel uses per source symbol, directly affecting the communication rate. The key rate constraint (\ref{eq:key-rate}) limits the secret key resource shared between the encoder and the legitimate decoder. The distortion constraints (\ref{eq:sem-distortion})--(\ref{eq:obs-distortion}) ensure fidelity, while the equivocation constraints {\color{blue}(\ref{ineq:sem-eqv})-(\ref{ineq:joint-eqv})} quantify the uncertainty of the eavesdropper about each source component and their joint version.

{\color{blue}
The individual equivocations allow to consider per-component secrecy. Additional joint equivocation constraint is necessary because the maximum of individual equivocations (semantic and observation), in general, does not lead to the maximum of the joint equivocation. Consider the following example. Let $S \sim \mathcal{B}(0.5)$, $U = S \oplus B$, where $B \sim \mathcal{B}(0.5)$. In this case, $H(S) = H(U) = 1$ and $H(S,U) = 2$. Assume $Z = S \oplus U$. Then, individual equivocations are at the maximum $H(S|Z) = 1$ and $H(U|Z) = 1$, while the joint $H(S,U|Z) = 1 \neq H(S|Z) + H(U|Z)$.
}

\begin{definition}[Achievable Region]
$\mathcal{R}$ is the set of all achievable $(r, R_k, D_s, D_u, \Delta_s, \Delta_u, \Delta_{su})$ and is referred to as the rate-distortion-equivocation region.
\end{definition}

\subsection{Background: Rate-Distortion Functions}
Our results, specifically the converse bound, rely on rate-distortion functions (RDFs) adapted to semantic communication scenarios. Consider the standard source coding problem, where the encoder $f : \mathcal{U}^k \to \mathcal{M}$ produces the message $M$ with rate $R$. The decoder $\hat{f} : \mathcal{M} \to \hat{\mathcal{U}}^k$ reconstructs $U^k$. The problem is to find the function $R_u(D_u) = \inf \{ R : (R,D_u) \text{ is achievable} \}$, where the tuple $(R,D_u)$ is achievable if there exists a code $(R,k)$ such that the distortion constraint (\ref{eq:obs-distortion}) holds. The following Lemma~\ref{lemma:classic-rdf} provides characterisation of the rate-distortion function in this case.

\begin{lemma}[Classic RDF \cite{elgamal2011book}]
\label{lemma:classic-rdf}
For a discrete memoryless source (DMS) $U$ with distortion constraint $D_u$, the rate-distortion function is:
\begin{equation}
R_u(D_u) = \inf_{\substack{p_{\hat{U}|U} \\ \EX d_u(U,\hat{U}) \leq D_u}} I(U;\hat{U}).
\end{equation}
\end{lemma}

Furthermore, assume that the encoder $f : \mathcal{S}^k \times \mathcal{U}^k \to \mathcal{M}$ produces the message $M$ for both $S^k$ and $U^k$ with rate $R$. The decoder $\hat{f} : \mathcal{M} \to (\hat{\mathcal{S}}^k, \hat{\mathcal{U}}^k)$ reconstructs $S^k$ and $U^k$. Now the problem is to find the function of two arguments, $R(D_s,D_u) = \inf \{ R : (R,D_s,D_u) \text{ is achievable} \}$. The tuple $(R,D_s,D_u)$ is achievable if there exists a code $(R,k)$ such that the distortion constraints (\ref{eq:sem-distortion}) and (\ref{eq:obs-distortion}) hold. Lemma~\ref{lemma:direct-rd} presents characterisation of the RDF in this case.

\begin{lemma}[Semantic RDF \cite{elgamal1982} for Case 2] \label{lemma:direct-rd}
For a semantic source $p_{S,U}$ with an encoder that has direct access to samples of both source components (Case 2), the rate-distortion function, under separate distortion constraints $D_s$ and $D_u$, is:
\begin{equation}
R_d(D_s, D_u) = \inf_{\substack{p_{\hat{S},\hat{U}|S,U} \\ \EX d_s(S,\hat{S}) \leq D_s, \\ \EX d_u(U,\hat{U}) \leq D_u}} I(S,U;\hat{S},\hat{U}),
\end{equation}
where \( \hat{S} \) and \( \hat{U} \) denote lossy reconstructions of the semantic and observation components.
\end{lemma}

When restricted to observation-only encoding, i.e., $f : \mathcal{U}^k \to \mathcal{M}$  (Case 1), one can derive an alternative RDF characterisation (Lemma~\ref{lemma:indirect-rd}) through the Markov chain \( S \rightarrow U \rightarrow (\hat{S},\hat{U}) \) and a modified distortion metric \cite{dobrushin1962, indirect1980}. 

\begin{lemma}[Semantic RDF \cite{liu2022indirect} for Case 1] \label{lemma:indirect-rd}
For a semantic source $p_{S,U}$ with encoder having access only to observation, the rate-distortion function becomes:
\begin{equation}
R_i(D_s, D_u) = \inf_{\substack{p_{\hat{S},\hat{U}|U} \\ \EX d_u(U,\hat{U}) \leq D_u, \\ \EX \hat{d}_s(U,\hat{S}) \leq D_s}} I(U;\hat{S},\hat{U}),
\end{equation}
where $\hat{d}_s(U, \hat{S}) = \sum_{s \in S} p_{S|U}(s|U) d_s(s,\hat{S})$ is the modified distortion metric.
\end{lemma}
This modified distortion metric guarantees the equivalence of two average distortions, i.e., $\EX \hat{d}_s(U, \hat{S}) = \EX d_s(S, \hat{S})$, simplifies problem to the classic source coding problem \cite{dobrushin1962}, and allows minimisation over \( p_{\hat{S},\hat{U}|U} \) rather than \( p_{\hat{S},\hat{U}|S,U} \), reducing the optimisation complexity.
\section{Converse Result} \label{sec:converse-result}

In this section, we establish a converse result on the fundamental limits of semantic communications over wiretap channels for the system model defined in Section \ref{sec:problem}.

\begin{theorem}[Converse]
\label{theorem:converse}
For a semantic source $p_{S,U}$ and a {\color{blue}general wiretap channel $p_{Y,Z|X}$} any achievable tuple $(r, R_k, D_s, D_u, \Delta_s, \Delta_u, \Delta_{su})$ must satisfy:
\ifCLASSOPTIONdraft
\begin{align}
    R(D_s,D_u) &\leq r I(X;Y) \label{ineq:converse-rate} \\
    \Delta_s &\leq H(S) - R(D_s) + R_k + r \max_{p_{\App}} {\color{blue} \left( I(\App;Y) - I(\App;Z) \right)} \label{ineq:converse-sem-eqv} \\
    \Delta_u &\leq H(U) - R(D_u) + R_k + r \max_{p_{\Bpp}} {\color{blue} \left( I(\Bpp;Y) - I(\Bpp;Z) \right)} \label{ineq:converse-obs-eqv} \\
    \Delta_{su} &\leq H(S,U) - R(D_s,D_u) + R_k + r \max_{p_{\ABpp}} {\color{blue} \left( I(\ABpp;Y) - I(\ABpp;Z) \right)}\label{ineq:converse-joint-eqv}
\end{align}
\else
\begin{align}
    R(D_s,D_u) &\leq r I(X;Y) \label{ineq:converse-rate} \\
    \Delta_s &\leq H(S) - R(D_s) + R_k
    \nonumber \\ &\quad + r \max_{p_{\App}} {\color{blue} \left( I(\App;Y) - I(\App;Z) \right)} \label{ineq:converse-sem-eqv} \\
    \Delta_u &\leq H(U) - R(D_u) + R_k \nonumber \\ &\quad + r \max_{p_{\Bpp}} {\color{blue} \left( I(\Bpp;Y) - I(\Bpp;Z) \right)} \label{ineq:converse-obs-eqv} \\
    \Delta_{su} &\leq H(S,U) - R(D_s,D_u) + R_k
    \nonumber \\ &\quad + r \max_{p_{\ABpp}} {\color{blue} \left( I(\ABpp;Y) - I(\ABpp;Z) \right)}\label{ineq:converse-joint-eqv}
\end{align}
\fi
where $\ABpp = (\App,\Bpp)$, ${\color{blue} |\mathcal{W}_1|, |\mathcal{W}_2|} \leq |\mathcal{X}|$, are auxiliary r.v.s with joint distribution $p_{\App,\Bpp,X,Y,Z} = p_{\App} p_{\Bpp|\App} p_{X|\Bpp} p_{Y,Z|X}$ in Case 1 and $p_{\App,\Bpp,X,Y,Z} = p_{\App,\Bpp} p_{X|\App,\Bpp} p_{Y,Z|X}$ in Case 2 encoding.
\end{theorem}
\begin{remark}
   Inequality (\ref{ineq:converse-rate}) implies inequalities with the marginal versions of RDFs, i.e., $R(D_s) \leq r I(X;Y)$ and $R(D_u) \leq r I(X;Y)$.
\end{remark}

\begin{remark}
\label{remark:converse}
Further improvement of the converse bound can be done with separate key rates for each source component.
\end{remark}

The converse result in Theorem~\ref{theorem:converse} has the following interpretation. Inequality (\ref{ineq:converse-rate}) bounds communication rate given the distortion constraints to the main channel capacity. The fundamental limit for secrecy is governed by remaining inequalities (\ref{ineq:converse-sem-eqv})-(\ref{ineq:converse-joint-eqv}). Generally, for this model, secrecy (equivocation) conditions consist of the following basic terms.
1) The level of compression of the semantic source, e.g., $H(S) - R_s(D_s)$ for the semantic part. This term naturally arises from the loss of entropy in source encoding and it also, however, contributes to the equivocation for legitimate receiver. The higher distortion for a specific component that we enforce, the higher equivocation can be achieved for this component.
2) $R_k$ -- rate of the secret key shared between the encoder and the legitimate decoder. The higher this rate, the better secrecy can be achieved for both components of the semantic source. The model allows to set this key rate to zero, thus removing source encryption. 
3) Secrecy capacity of the wiretap channel, e.g, $ \max_{p_{\App}} {\color{blue} \left( I(\App;Y) - I(\App;Z) \right) }$ for the semantic part and $\max_{p_{\Bpp}} {\color{blue} \left( I(\Bpp;Y) - I(\Bpp;Z) \right) }$ for the observed part. These terms are scaled by $r$ and are determined by the statistics of the main and the eavesdropper channels. The total available secrecy capacity of the wiretap channel is distributed to secure semantic $\max_{p_{\App}} {\color{blue} \left( I(\App;Y) - I(\App;Z) \right) }$ and observed $\max_{p_{\Bpp}} {\color{blue} \left( I(\Bpp;Y) - I(\Bpp;Z) \right) }$ parts of the source. The split of the secrecy capacity is determined by $\App$ and $\Bpp$ r.v.s.
\begin{remark}
   For the degraded wiretap channel $p_{Y,Z|X} = p_{Y|X}p_{Z|Y}$, every secrecy capacity term can be written as $r \max_{p_X} I(X;Y|Z)$.
\end{remark}

\subsection{Comparison to Known Results}
In this subsection, we show that our model and the converse result generalise some of the previously established results \cite{yamamoto1983, yamamoto1997, guo2022}.

\subsubsection{Secure lossy source coding of semantic sources}
The basic model of secure semantic source coding is presented in \cite{yamamoto1983}, where the source modelling is identical to our definition of the semantic source. The model assumes the lossy encoder, producing message $W \in [0,...,M - 1]$ with rate $R = \tfrac{1}{k} \log M$, the legitimate decoder, and the eavesdropper. The communication link between the encoder and the decoder is assumed to be noise-free, and no secret key is shared between the encoder and the legitimate decoder. Thus, secrecy is achieved only due to lossy compression. The fidelity constraint is imposed only to the observed part of the source, and the secrecy constraint is considered only for the semantic part of the source. In such case, the following rate-distortion-equivocation tradeoff is valid.
\begin{theorem}[{\hspace{1sp}\cite[Theorem 3]{yamamoto1983}}]
The following tight bound is valid for achievable tuple $(R,D_u,\Delta_s)$
\begin{align}
    \EX d_u(U,\hat{U}) &\leq D_u, \\
    R &\geq I(S,U;\hat{U}), \\
    \Delta_s &\leq H(S|\hat{U})
\end{align}
\end{theorem}
This bound is valid for Case 1 and Case 2 encoding. In Case 1 the term $I(S,U;\hat{U})$ simplifies to $I(U;\hat{U})$, and the term $H(S|\hat{U})$ can be substituted with $H(U|\hat{U})$.

\subsubsection{Secure source coding of semantic sources with secret key} Later, the extension of the problem \cite{yamamoto1983} to one with secret key shared between encoder and decoder is studied in \cite{guo2022}, where the authors derived a tight bound for the rate-distortion-equivocation region. Their model also consists of semantic source, source encoder that produces message $W \in [0,...,M - 1]$ with rate $R = \tfrac{1}{k} \log M$, noise-free link, the legitimate decoder, and the eavesdropper. However, to enable secrecy, the rate-limited with $R_k$ secret key is shared between the encoder and the legitimate decoder.
In \cite{guo2022}, the authors establish the following region for secure semantic source coding.
\begin{theorem}[{\hspace{1sp}\cite[Theorem 2]{guo2022}}]
\label{theorem:guo}
    For a semantic source $p_{S,U}$ in the secure source coding setting under Case 1 encoding, the following tight bound is valid for tuple $(R, R_k, \Delta_s, D_s, D_u)$:
    \ifCLASSOPTIONdraft
    \begin{align}
        R &\geq R_i(D_s,D_u), \\
        \Delta_s &\leq R_k +  H(S) - R(D_s). \label{eq:guo-inf-leak}
    \end{align}
    \else
    \begin{align}
        &R \geq R_i(D_s,D_u), \
        \Delta_s \leq R_k +  H(S) - R(D_s). \label{eq:guo-inf-leak}
    \end{align}
    \fi
\end{theorem}
The authors of Theorem~\ref{theorem:guo} originally use the information leakage secrecy metric, i.e., $\tfrac{1}{k} I(S^k;X)$. To better compare with our converse result, we substitute information leakage with equivocation metric, $\tfrac{1}{k} H(S^k|X) = \tfrac{1}{k} H(S^k) - \tfrac{1}{k} I(S^k;X)$. We also explicitly use the rate-distortion function from Lemma~\ref{lemma:indirect-rd}. This is because the inequalities (4), (6), and (7) in \cite[Theorem 2]{guo2022} are equivalent to $R \geq R_i(D_s,D_u)$. With these substitutions, it can be seen that the region in Theorem~\ref{theorem:guo} is the special case of the region in Theorem~\ref{theorem:converse}, if one discards the noisy channel ($X=Y=Z$), sets $n=1$, and discards the secrecy constraint for observation, i.e., (\ref{ineq:obs-eqv}) and (\ref{ineq:converse-obs-eqv}), in our problem formulation. 

\subsubsection{Secure lossy JSCC of classic sources} In addition, our converse region also reduces to the secure lossy source-channel coding problem of a classic source, solved by \cite{yamamoto1997}. The model in this case is the one with the classic source model and single distortion and secrecy constraints. The author \cite{yamamoto1997} presents the following region, which is also the special case of our region.
\begin{theorem}[{\hspace{1sp}\cite[Theorem 1]{yamamoto1997}}]
The following tight bound is valid for the system model with source $p_{U}$ transmitted via the degraded wiretap channel $p_{Y|X}p_{Z|Y}$, with a secret key rate $R_k$ shared between the encoder and legitimate decoder, given fidelity $D_u$, and secrecy $\Delta_u$.
\ifCLASSOPTIONdraft
\begin{align}
    R(D_u) &\leq r \cdot I(X;Y) \\
    \Delta_u &\leq R_k + r \cdot \left[ I(X;Y) - I(X;Z) \right] + H(U) - R(D_u)
\end{align}
\else
\begin{align}
    &R(D_u) \leq r \cdot I(X;Y) \\
    &\Delta_u {\leq} R_k {+} r \left[ I(X;Y) {-} I(X;Z) \right] {+} H(U) {-} R(D_u)
\end{align}
\fi
\end{theorem}

The equivalent model in this case will be one with Case 1 encoding and without semantic distortion (\ref{eq:sem-distortion}) and equivocation constraints (\ref{ineq:sem-eqv}), (\ref{ineq:joint-eqv}).

\subsection{Proof of Theorem~\ref{theorem:converse}}
\label{subsec:converse-proof}
To prove the converse, firstly, we provide established inequities that bound RDFs and channel single-letter input-output mutual information to multi-letter mutual information terms. Then, we rely on the presented inequalities to derive the rate-distortion region. To show the converse region for equivocation, we bound the secret key rate to single-letter terms and RDFs, following the idea of \cite{yamamoto1997}. The derivation in part is based on \cite[Lemma 17.12]{csiszar2011book}.

\subsubsection{Supplementary Results}
The following inequalities hold for a memoryless semantic source $p_{SU}$, given $\epsilon > 0$,
\ifCLASSOPTIONdraft
\begin{align}
    \label{ineq:sem-rdf}
	\frac{1}{k} I(S^k; \hat{S}^k) &\geq^{(a)} \frac{1}{k} \sum_{i=1}^{k} I(S_i; \hat{S}_i) \geq^{(b)} R_s(D_s + \epsilon), \\
    \label{ineq:obs-rdf}
	\frac{1}{k} I(U^k; \hat{U}^k) &\geq^{(a)} \frac{1}{k} \sum_{i=1}^{k} I(U_i; \hat{U}_i) \geq^{(b)} R_u(D_u + \epsilon), \\
    \label{ineq:joint-rdf}
    \frac{1}{k} I(S^k, U^k; \hat{S}^k, \hat{U}^k) &\geq^{(a)} \frac{1}{k} \sum_{i=1}^{k} I(S_i, U_i;\hat{S}_i, \hat{U}_i)
    \geq^{(b)} R(D_s + \epsilon, D_u + \epsilon),
\end{align}
\else
\begin{align}
    \label{ineq:sem-rdf}
	&\frac{1}{k} I(S^k; \hat{S}^k) \overset{(a)}{\geq} \frac{1}{k} \sum_{i=1}^{k} I(S_i; \hat{S}_i) \overset{(b)}{\geq} R_s(D_s + \epsilon), \\
    \label{ineq:obs-rdf}
	&\frac{1}{k} I(U^k; \hat{U}^k) \overset{(a)}{\geq} \frac{1}{k} \sum_{i=1}^{k} I(U_i; \hat{U}_i) \overset{(b)}{\geq} R_u(D_u + \epsilon), \\
    \label{ineq:joint-rdf}
    &\frac{1}{k} I(S^k, U^k; \hat{S}^k, \hat{U}^k) \geq^{(a)} \frac{1}{k} \sum_{i=1}^{k} I(S_i, U_i;\hat{S}_i, \hat{U}_i) \nonumber \\
    &\geq^{(b)} R(D_s + \epsilon, D_u + \epsilon),
\end{align}
\fi
where (a) is due to the memoryless property of the source and \cite[Theorem 6.1]{polyanskiy2025book}, and (b) because RDF is a convex and non-increasing function.

In addition, for the memoryless channel $p_{Y^n|X^n} = \prod_{i=1}^{n} p_{Y_i|X_i}$, the following inequalities are valid \cite{elgamal2011book, polyanskiy2025book},
\begin{equation} 
    \label{ineq:ch-ineqs}
	I(X^n; Y^n) \leq \sum_{i=1}^{n} I(X_i; Y_i) \leq n I(X;Y).
\end{equation}

\subsubsection{Rate-Distortion Region}
We now prove the converse bound for rate-distortion (\ref{ineq:converse-rate}) as,
\ifCLASSOPTIONdraft
\begin{align}
    \label{ineq:rd-region-proof-line-1}
	R(D_s + \epsilon, D_u + \epsilon)
    &\leq^{(\ref{ineq:joint-rdf})}
    \frac{1}{k} \sum_{i=1}^k I(S_i, U_i;\hat{S}_i, \hat{U}_i)
    \leq^{(a)} \frac{1}{k} \sum_{i=1}^n I(X_i;Y_i) \\
    &\leq^{(\ref{ineq:ch-ineqs})}
    \frac{n}{k} I(X;Y)
    \leq^{ (\ref{ineq:rate-cond}) }
    (r + \epsilon) I(X;Y),
    \label{ineq:rd-region-proof-line-2}
\end{align}
\else
\begin{align}
	&R(D_s + \epsilon, D_u + \epsilon)
    \leq^{(\ref{ineq:joint-rdf})}
    \frac{1}{k} \sum_{i=1}^k I(S_i, U_i;\hat{S}_i, \hat{U}_i) \nonumber \\
    &\overset{(a)}{\leq} \frac{1}{k} \sum_{i=1}^n I(X_i;Y_i)
    \overset{(\ref{ineq:ch-ineqs})}{\leq} \frac{n}{k} I(X;Y)
    \overset{(\ref{ineq:rate-cond})}{\leq} (r + \epsilon) I(X;Y),
    \label{ineq:rd-region-proof-line-2}
\end{align}
\fi
where $\epsilon > 0$, and (a) is due to the data processing inequality (DPI).

\subsubsection{Equivocation Region}
The proof of the bound for semantic equivocation (\ref{ineq:converse-sem-eqv}) is as follows.
\ifCLASSOPTIONdraft
\begin{align} \label{eq:sem-eqv-first-line}
    k(R_k + \epsilon) &\geq^{(\ref{eq:key-rate})} \log |\mathcal{K}| \geq H(K) \geq H(K|Y^n) \geq H(K|Y^n) - H(K|Y^n, \hat{S}^k) \\
    &=  H(K, Y^n) - H(Y^n) - H(K, Y^n, \hat{S}^k) + H(Y^n, \hat{S}^k) \\
    &= H(\hat{S}^k|Y^n) - H(\hat{S}^k|Y^n, K) =^{(a)} H(\hat{S}^k|Y^n) \\
    &\geq^{(\ref{ineq:sem-eqv})} H(\hat{S}^k|Y^n) - \bigl( H(S^k|Z^n) - k (\Delta_s - \epsilon) \bigr) \\
    &= \bigl( H(\hat{S}^k, Y^n) - H(Y^n) \bigr) - \bigl( H(S^k, Z^n) - H(Z^n) \bigr) - k (\Delta_s - \epsilon) \\
    &= I(S^k;Z^n) - I(\hat{S}^k;Y^n) + H(\hat{S}^k|S^k) - H(S^k) + I(\hat{S}^k;S^k) + k (\Delta_s - \epsilon), \label{eq:sem-eqv-three-terms}
\end{align}
\else
\begin{align} \label{eq:sem-eqv-first-line}
    &k(R_k + \epsilon) \geq^{(\ref{eq:key-rate})} \log |\mathcal{K}| \geq H(K) \geq H(K|Y^n) 
    \\ &\geq H(K|Y^n) - H(K|Y^n, \hat{S}^k) =  H(K, Y^n) \nonumber \\
    &\quad - H(Y^n) - H(K, Y^n, \hat{S}^k) + H(Y^n, \hat{S}^k) \\
    &= H(\hat{S}^k|Y^n) - H(\hat{S}^k|Y^n, K) =^{(a)} H(\hat{S}^k|Y^n) \\
    &\geq^{(\ref{ineq:sem-eqv})} H(\hat{S}^k|Y^n) - \bigl( H(S^k|Z^n) - k (\Delta_s - \epsilon) \bigr) \\
    &= \bigl( H(\hat{S}^k, Y^n) - H(Y^n) \bigr) - \bigl( H(S^k, Z^n) - H(Z^n) \bigr) \nonumber \\
    &\quad + k (\Delta_s - \epsilon) = I(S^k;Z^n) - I(\hat{S}^k;Y^n) \\
    &\quad {+} H(\hat{S}^k|S^k) {-} H(S^k) {+} I(\hat{S}^k;S^k) + k (\Delta_s {-} \epsilon), \label{eq:sem-eqv-three-terms}
\end{align}
\fi
where (a) because $\hat{S}^k$ can be fully reconstructed with channel output $Y^n$ and secret key $K$. 
The first three terms of (\ref{eq:sem-eqv-three-terms}) can be rewritten as follows:
\ifCLASSOPTIONdraft
\begin{align}
    &I(S^k; Z^n) - I(\hat{S}^k; Y^n) + H(\hat{S}^k|S^k) = I(S^k; Z^n) -  H(Y^n) + H(Y^n|\hat{S}^k) + H(\hat{S}^k|S^k) \\
    &\geq I(S^k; Z^n) {-}  H(Y^n) {+} H(Y^n|\hat{S}^k, S^k) {+} H(\hat{S}^k|S^k)
    = I(S^k; Z^n) {-} H(Y^n) {+} H(Y^n, \hat{S}^k|S^k) \\
    &\geq I(S^k;Z^n) - I(S^k;Y^n)
    =^{(a)} n I(\App;Z|\ABcc) - n I(\App;Y|\ABcc) \\
    &{\color{blue} = - n \EX_Q \left( I(\App;Y|\ABcc {=} q) {-} I(\App;Z|\ABcc {=} q) \right) \geq - n \max_q \left( I(\App;Y|\ABcc {=} q) {-} I(\App;Z|\ABcc {=} q) \right)} \\
    &{\color{blue} \geq - n \max_q \max_{p_{\App}} \left( I(\App;Y|\ABcc {=} q) {-} I(\App;Z|\ABcc {=} q) \right) = - n \max_{p_{\App}} \left( I(\App;Y) {-} I(\App;Z) \right)}
    \label{eq:sem-eqv-secrecy-capacity}
\end{align}
\else
\begin{align}
    &I(S^k; Z^n) - I(\hat{S}^k; Y^n) + H(\hat{S}^k|S^k) \\
    &= I(S^k; Z^n) {-}  H(Y^n) {+} H(Y^n|\hat{S}^k) {+} H(\hat{S}^k|S^k) \\
    &{\geq} I(S^k {;} Z^n) {-}  H(Y^n) {+} H(Y^n|\hat{S}^k S^k) {+} H(\hat{S}^k|S^k)
    \\ &= I(S^k; Z^n) {-} H(Y^n) {+} H(Y^n, \hat{S}^k|S^k) \\
    &\geq I(S^k;Z^n) - I(S^k;Y^n)
    \\ &=^{(a)} n I(\App;Z|\ABcc) - n I(\App;Y|\ABcc)
    &
    \\ &= - n \EX_Q \left( I(\App;Y|\ABcc {=} q) {-} I(\App;Z|\ABcc {=} q) \right) \\
    &\geq - n \max_q \left( I(\App;Y|\ABcc {=} q) {-} I(\App;Z|\ABcc {=} q) \right)
    \\ &{\geq} {-} n \max_q \max_{p_{\App}} \left( I(\App{;}Y|\ABcc {=} q) {-} I(\App {;} Z|\ABcc {=} q) \right) \\
    &= - n \max_{p_{\App}} \left( I(\App;Y) {-} I(\App;Z) \right)
    \label{eq:sem-eqv-secrecy-capacity}
\end{align}
\fi
where (a) is due to \cite[Lemma 17.12]{csiszar2011book} with $\ABcc = (J,Y^{J-1},Z_{J+1}^{n})$, $\App = (\ABcc,S^k)$, $Y = Y_J$, $Z = Z_J$, and $J$ is a uniform r.v. on $\{1,...,n\}$. {\color{blue} Note that given fixed $Q=q$ the following Markov chain is valid $\App \to X \to (Y,Z)$ because the channel is memoryless}. Replacing in (\ref{eq:sem-eqv-three-terms}) we have
\ifCLASSOPTIONdraft
\begin{align}
    &k(R_k + \epsilon) \geq^{(a)} - n \max_{p_{\App}} \left[ I(\App;Y) - I(\App;Z) \right] - k H(S) + k R_s(D_s) + k (\Delta_s - \epsilon).
\end{align}
\else
\begin{align}
    &k(R_k + \epsilon) \geq^{(a)} - n \max_{p_{\App}} \left[ I(\App;Y) - I(\App;Z) \right] \nonumber \\
    \quad &- k H(S) + k R_s(D_s) + k (\Delta_s - \epsilon).
\end{align}
\fi
where (a) is due to (\ref{eq:sem-eqv-secrecy-capacity}), (\ref{ineq:sem-rdf}), and the fact that $S^k$ is i.i.d.

Finally, we divide all terms in (\ref{eq:sem-eqv-three-terms}) by $k$, rearrange them, and use (\ref{ineq:rate-cond}), to obtain
\ifCLASSOPTIONdraft
\begin{equation}
    \Delta_s - 2\epsilon \leq R_k + (r + \epsilon) \cdot \max_{p_{\App}} \left[ I(\App;Y) - I(\App;Z) \right] + H(S) - R_s(D_s). \label{eq:sem-eqv-last-line}
\end{equation}
\else
\begin{align}
    &\Delta_s - 2\epsilon \leq R_k + (r + \epsilon) \cdot \max_{p_{\App}} \left[ I(\App;Y) - I(\App;Z) \right] \nonumber \\
    &\quad + H(S) - R_s(D_s). \label{eq:sem-eqv-last-line}
\end{align}
\fi

To prove the bound for observation equivocation (\ref{ineq:converse-obs-eqv}), we follow the same steps (\ref{eq:sem-eqv-first-line})-(\ref{eq:sem-eqv-last-line}) but with $(U^k,\hat{U}^k)$ instead of $(S^k,\hat{S}^k)$, constraint (\ref{ineq:obs-eqv}) instead of (\ref{ineq:sem-eqv}), (\ref{ineq:obs-rdf}) instead of (\ref{ineq:sem-rdf}), and $\Bpp = (Q,U^k)$ instead of $\App$. {\color{blue} We note that in Case 1 encoding we assume $\App \to \Bpp \to X \to (Y,Z)$ Markov chain to hold. This is due to $S^k \to U^k \to X^n \to (Y^n,Z^n)$ in Case 1 encoding.}
Similarity, we obtain the bound for joint equivocation (\ref{ineq:converse-joint-eqv}), using $(S^k, U^k, \hat{S}^k, \hat{U}^k)$, constraint (\ref{ineq:joint-eqv}), and inequality (\ref{ineq:joint-rdf}). For joint equivocation, we also set the auxiliary r.v. to $\ABpp = (S^k,U^k,\ABcc) = (\App,\Bpp)$ when applying \cite[Lemma 17.12]{csiszar2011book}. {\color{blue}The proof of cardinality bounds for $(\App,\Bpp)$ is presented in Appendix~\ref{appendix:cardinality}.}
Letting $\epsilon \to 0$ completes the converse proof.

\section{Achievability Scheme}
\label{sec:ach-result}
\begin{theorem}[Achievability]
\label{theorem:direct}
For a semantic source $p_{S,U}$, a {\color{blue} general wiretap channel $p_{Y,Z|X}$}, a tuple $(r, {\color{blue} R_k}, D_s, D_u, \Delta_s, \Delta_u, \Delta_{su})$ is achievable if there exist auxiliary r.v.s $\A = (\Sc,\Sp)$, $\B = (\Uc,\Up)$, {\color{blue} which are arbitrary jointly distributed in Case 2 encoding and given Markov chain $S \to U \to (\A, \B)$ in Case 1 encoding, $\ABcc =(\Acc,\Bcc)$, $\ABpp = (\App,\Bpp)$, so that $(\ABcc,\ABpp) \to X \to (Y,Z)$,} and functions {\color{blue} $\tilde{\mathcal{S}} : \mathcal{\A} \rightarrow \hat{\mathcal{S}}$ and $\tilde{\mathcal{U}} : \mathcal{\A}_0 \times \mathcal{\B} \rightarrow \hat{\mathcal{U}}$}, such that the following inequalities hold:
\ifCLASSOPTIONdraft
\begin{align}
\label{ineq:ach-rate-first}
&D_s \leq \EX d_s(S, \tilde{S} (\A)), \\
&D_u \leq \EX d_u(U, \tilde{U} (\Sc,\B)), \\
&I(\Sc;V) < r \cdot I(\Acc;Y), \\
&{\color{blue} I(\A;V)} < r \cdot I(\Acc,\App;Y), \\
&I(\B;V|\Sc) < r \cdot I(\Bcc,\Bpp;Y|\Acc), \label{ineq:ach-rate-last}\\
& \Delta_{s} \leq [H(S) - I(\A;V) {\color{blue} \ - I(S;\B|\A)]  + R_{k1}} + r [ I(\App;Y|\Acc) - I(\App;Z|\Acc) ], \\
&\Delta_{u} \leq [H(U) - I(\B;V|\Sc) {\color{blue} \ - I(U;\Sp|\Sc,\B)] + R_{k} } + r [ I(\Bpp;Y|\ABcc) {-} I(\Bpp;Z|\ABcc) ], \\
&\Delta_{su} \leq [H(S,U) - I(\A;V) - I(\B;V|\Sc)] {\color{blue} \ + R_{k}}  + r [ I(\App;Y|\Acc) + I(\Bpp;Y|\ABcc) - I(\ABpp;Z|\ABcc) ],
\end{align}
\else
\begin{align}
\label{ineq:ach-rate-first}
&D_s \leq \EX d_s(S, \tilde{S} (\A)), \\
&D_u \leq \EX d_u(U, \tilde{U} (\Sc,\B)), \\
&I(\Sc;V) < r \cdot I(\Acc;Y), \\
&{\color{blue} I(\A;V)} < r \cdot I(\Acc,\App;Y), \\
&I(\B;V|\Sc) < r \cdot I(\Bcc,\Bpp;Y|\Acc), \label{ineq:ach-rate-last}\\
& \Delta_{s} \leq [H(S) - I(\A;V) {\color{blue} \ - I(S;\B|\A)]  + R_{k1}} \nonumber \\
&\quad + r [ I(\App;Y|\Acc) - I(\App;Z|\Acc) ], \\
&\Delta_{u} \leq [H(U) - I(\B;V|\Sc) {\color{blue} \ - I(U;\Sp|\Sc,\B)] + R_{k} } \nonumber \\
&\quad + r [ I(\Bpp;Y|\ABcc) {-} I(\Bpp;Z|\ABcc) ], \\
&\Delta_{su} \leq [H(S,U) - I(\A;V) - I(\B;V|\Sc)] {\color{blue} \ + R_{k}}  \nonumber \\
&\quad + r [ I(\App;Y|\Acc) {+} I(\Bpp;Y|\ABcc) {-} I(\ABpp;Z|\ABcc) ],
\end{align}
\fi
where in Case 1 encoder we define $V \doteq U$, and in Case 2 we define $V \doteq (S,U)$, {\color{blue} given $R_k = R_{k1} + R_{k2}$, $R_{k1} \leq I(\Sc;V)$ and $R_{k2} \leq I(\Uc;V|\Sc)$}.
\end{theorem}
Equations (\ref{ineq:ach-rate-first})-(\ref{ineq:ach-rate-last}) correspond to {\color{blue} constraint} (\ref{ineq:converse-rate}) in converse bound and define an achievable region of semantic source compression. The rest defines the secrecy region, where equivocation consists of {\color{blue} three} basic parts, as in converse bound. The part due to loss in source encoding {\color{blue}, e.g. $[H(S) - I(\A;V) - I(S;\B|\A)]$ , the part due to secret key rate $R_{k1}$ or $R_k$,} and the part due to wiretap channel coding, e.g {\color{blue}$r [I(\App;Y|\Acc) - I(\App;Z|\Acc)]$}. {\color{blue} One can see that if $H(S) - I(\A;V) - I(S;\B|\A) < r [I(\App;Y|\Acc) - I(\App;Z|\Acc)]$ then the secrecy for the semantic part is dominated by the secrecy capacity. Besides, if $R_{k1} > r [I(\App;Y|\Acc) - I(\App;Z|\Acc)]$ then the encryption provides more equivocation to semantics than the secrecy capacity.}

\subsection{Proof of Theorem~\ref{theorem:direct}}
\label{appendix:direct-proof}
{\color{blue} This subsection presents the proof of the achievability bound (Theorem~\ref{theorem:direct}). The scheme is based on a four-layer stochastic superposition code and uses auxiliary random variables $\A$ and $\B$ for the source coding stage and $\ABcc$ and $\ABpp$ for the channel coding stage, each consisting of two layers.

For the semantic component $S^k$, the auxiliary variable $\A$ generates a lossy source codebook and is split into a partially encrypted layer $\Sc$ and a wiretap-protected layer $\Sp$. Similarly, the observation component $U^k$ is represented by $\B$, which is split into $\Uc$ and $\Up$. This two-layer structure together with the secret key rate enables explicit control over which portions of the source description are public, encrypted and which are protected via wiretap coding. To avoid redundancy, the encrypted and wiretap-protected portions are supposed not to overlap. The use of separate auxiliaries $\A$ and $\B$ also allows the distortion requirements for $S^k$ and $U^k$ to be handled independently.

In the channel coding stage, two private layers $\App$ and $\Bpp$ are introduced and protected using wiretap coding, enabling separate equivocation control for the semantic and observation components. A single private layer, as in the classical wiretap model, would generally permit control of only one secrecy metric (e.g., joint equivocation). The corresponding public layers $\Acc$ and $\Bcc$ facilitate the mapping from source indices to channel codewords via the mapping $g$ defined in our scheme.

The overall construction follows an operational separation approach, where source and channel encoders are optimised jointly but operate as distinct modules~\cite{villard2014}. Random binning is used to provide additional secrecy against eavesdropping.

We first describe the source codebook generation and encoding procedure, followed by the channel codebook construction and encoding strategy. The channel encoder uses stochastic encoding to conceal the two private message components. The source and channel stages are connected through the mapping $g$, which translates source coding indices into channel coding indices. Finally, we derive conditions under which encoding and decoding succeed and show that the proposed scheme achieves the required distortion and equivocation levels for the semantic and observation components under the stated single-letter constraints.
}
\subsubsection{Codebook Generation}

{\color{blue} We use two stand-alone codebooks for source and channel encoding stages. The source codebook consists of sequences $\sck,\spk,\uck,\upk$. The channel codebook consists of sequences $\qsn,\wsn,\qun,\wun$, where $\ws$ and $\wu$ are random indices for the wiretap coding. {\color{teal} We refer to the rate of some index $i \in [1,...,2^{k \RSc}]$ as $\RSc$. For example, we have source codebook rates $\RSc, \RSp, \RUc, \RUp$ and the channel codebook rates $\RAcc, \RApp, \RBcc, \RBpp$, including wiretap coding indices rates $\Rms, \Rmu$.} Fig.~\ref{fig:codebooks-pic} illustrates these codebooks, and the generation process is explained in detail in Appendix~\ref{appendix:codebook}.
}

\ifCLASSOPTIONdraft
\begin{figure}
\centering
\hfill
\subfigure[Source Codebook]{\includegraphics[width=0.485\linewidth,draft=False]{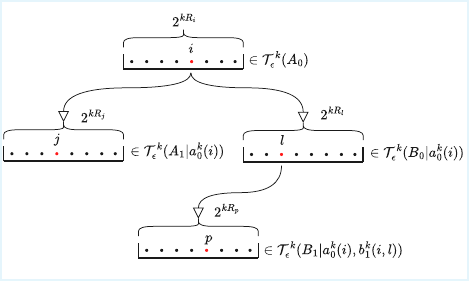}
\label{subfig:src-codebook}}
\hfill
\subfigure[Channel Codebook]{\includegraphics[width=0.485\linewidth,draft=False]{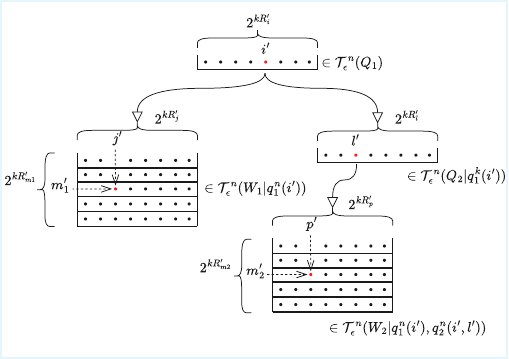}
\label{subfig:ch-codebook}}

\hfill
\caption{Codebook design. Indices $\ws$ and $\wu$ are picked at random from a uniform distribution.}
\label{fig:codebooks-pic}
\end{figure}
\else
\begin{figure}
\centering
\subfigure[Source Codebook]{\includegraphics[width=0.475\linewidth,draft=False]{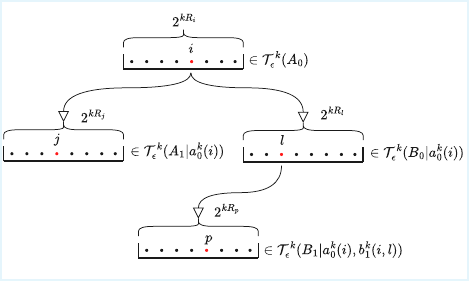}
\label{subfig:src-codebook}}
\hfill
\subfigure[Channel Codebook]{\includegraphics[width=0.475\linewidth,draft=False]{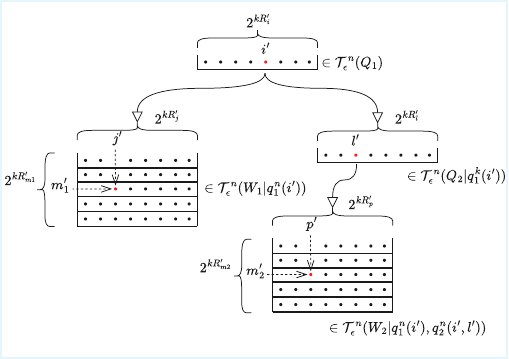}
\label{subfig:ch-codebook}}

\hfill
\caption{Codebook design. Indices $\ws$ and $\wu$ are picked at random from a uniform distribution.}
\label{fig:codebooks-pic}
\end{figure}
\fi

\subsubsection{Source Encoding Process}
\ifCLASSOPTIONdraft  
\begin{figure}
\centering
\hfill
\subfigure[Source Encoding Stage]{\ifCLASSOPTIONdraft 
\begin{tikzpicture}
\graph [] {
    a/"$\sck$"[at={(0,0)}];
    b/"$\uck$"[at={(2,1)}];
    c/"$\upk$"[at={(4,2)}];
    d/"$\spk$"[at={(2,2)}];
    a -> b -> c;
    a -> d;
  };
\end{tikzpicture}
\else
\begin{tikzpicture}
\graph [] {
    a/"$\sck$"[at={(0.0,0.0)}];
    b/"$\uck$"[at={(1.3,1)}];
    c/"$\upk$"[at={(1.3,1)}];
    d/"$\spk$"[at={(0,2)}];
    a -> b -> c;
    a -> d;
  };
\end{tikzpicture}
\fi
\label{subfig:src-encoding}}
\hfill
\subfigure[Channel Encoding Stage]{\ifCLASSOPTIONdraft
\begin{tikzpicture}
\graph [] {
    a/"$\qsn$"[at={(0,0)}];
    b/"$\qun$"[at={(2,1)}];
    c/"$\wun$"[at={(5,2)}];
    d/"$\wsn$"[at={(2.5,2)}];
    a -> b -> c;
    a -> d;
  };
\end{tikzpicture}
\else
\begin{tikzpicture}
\graph [] {
    a/"$\qsn$"[at={(0,0)}];
    b/"$\qun$"[at={(2.3,1)}];
    c/"$\wun$"[at={(2.3,1)}];
    d/"$\wsn$"[at={(0,2)}];
    a -> b -> c;
    a -> d;
  };
\end{tikzpicture}
\fi
\label{subfig:ch-encoding}}
\hfill
\caption{Superposition coding layouts}
\label{fig:encoding-graph}
\end{figure}
\else
\begin{figure}
\centering
\subfigure[Source Encoding Stage]{\small \ifCLASSOPTIONdraft 
\begin{tikzpicture}
\graph [] {
    a/"$\sck$"[at={(0,0)}];
    b/"$\uck$"[at={(2,1)}];
    c/"$\upk$"[at={(4,2)}];
    d/"$\spk$"[at={(2,2)}];
    a -> b -> c;
    a -> d;
  };
\end{tikzpicture}
\else
\begin{tikzpicture}
\graph [] {
    a/"$\sck$"[at={(0.0,0.0)}];
    b/"$\uck$"[at={(1.3,1)}];
    c/"$\upk$"[at={(1.3,1)}];
    d/"$\spk$"[at={(0,2)}];
    a -> b -> c;
    a -> d;
  };
\end{tikzpicture}
\fi
\label{subfig:src-encoding}}
\subfigure[Channel Encoding Stage]{\small \ifCLASSOPTIONdraft
\begin{tikzpicture}
\graph [] {
    a/"$\qsn$"[at={(0,0)}];
    b/"$\qun$"[at={(2,1)}];
    c/"$\wun$"[at={(5,2)}];
    d/"$\wsn$"[at={(2.5,2)}];
    a -> b -> c;
    a -> d;
  };
\end{tikzpicture}
\else
\begin{tikzpicture}
\graph [] {
    a/"$\qsn$"[at={(0,0)}];
    b/"$\qun$"[at={(2.3,1)}];
    c/"$\wun$"[at={(2.3,1)}];
    d/"$\wsn$"[at={(0,2)}];
    a -> b -> c;
    a -> d;
  };
\end{tikzpicture}
\fi
\label{subfig:ch-encoding}}
\caption{Superposition coding layouts}
\label{fig:encoding-graph}
\end{figure}
\fi

The encoding process depends on the encoder case as described in Section~\ref{sec:problem}. Let $v^k \doteq (s^k,u^k)$ and $V \doteq (S,U)$ in Case 2 (encoder has access to both semantic $s^k$ and observed $u^k$ parts of the semantic source), and $v^k \doteq u^k$ and $V \doteq U$ in Case 1 (encoder has access only to the observation $u^k$). The encoding is based on joint typicality, where the encoder finds first sequence such that:
\ifCLASSOPTIONdraft
\begin{align}
    (\sck, v^k) &\in \T^k(\Sc, V), \quad\quad\qquad (\spk, v^k) \in \T^k(\Sp, V | \sck), \\
    (\uck, v^k) &\in \T^k(\Uc, V|\sck), \quad (\upk, v^k) \in \T^k(\Up, V | \sck,\uck).
\end{align}
\else
\begin{align}
    &(\sck, v^k) \in \T^k(\Sc, V),
    \\&(\spk, v^k) \in \T^k(\Sp, V | \sck),
    \\&(\uck, v^k) \in \T^k(\Uc, V|\sck), \\&(\upk, v^k) \in \T^k(\Up, V | \sck,\uck).
\end{align}
\fi

Fig.~\ref{subfig:src-encoding} shows the sequence selection order. The encoding will be successful {\color{blue} if there will be at least one jointly typical sequence to pick from which holds} if rates satisfy \cite[Covering Lemma 3.3]{elgamal2011book}:
\ifCLASSOPTIONdraft
\begin{align}
\label{ineq:rate-Ri}
    \RSc &> I(\Sc; V) + \delta_{\color{blue} k}, \qquad
    \RSp > I(\Sp; V | \Sc) + \delta_{\color{blue} k},\\
\label{ineq:rate-Rp}
    \RUc &> I(\Uc; V | \Sc) + \delta_{\color{blue} k}, \
    \RUp > I(\Up; V | \Sc,\Uc) + \delta_{\color{blue} k},
\end{align}
\else
\begin{align}
\label{ineq:rate-Ri}
    &\RSc {>} I(\Sc; V) {+} \delta_{\color{blue} k}, \quad
    \RSp {>} I(\Sp; V | \Sc) {+} \delta_{\color{blue} k},\\
\label{ineq:rate-Rp}
    &\RUc {>} I(\Uc; V | \Sc) {+} \delta_{\color{blue} k}, 
    \RUp {>} I(\Up; V | \Sc,\Uc) {+} \delta_{\color{blue} k},
\end{align}
\fi
where $\delta_{\color{blue} k} \to 0$ with $k \to \infty$.

\subsubsection{\color{blue} Encryption}
{\color{blue} 
Let secret key $K$ be a uniform r.v.  independent of other r.v.s such that $\frac{1}{k} \log |\mathcal{K}| = R_k$. Consider {\color{teal} partition} of $K$ into independent $K_1$ and $K_2$ such that {\color{teal} $\mathcal{K} = \mathcal{K}_1 \times \mathcal{K}_2$ and} $R_{k1} + R_{k2} = R_k$ {\color{teal} so that} $R_{k1} \leq I(\Sc;V) \leq \RSc$, $R_{k2} \leq I(\Uc;V|\Sc) \leq \RUc$, where $R_{k1} = \frac{1}{k} \log |\mathcal{K}_1|$ and $R_{k2} = \frac{1}{k} \log |\mathcal{K}_2|$. Then, we encrypt indices $i$ and $l$ from the source encoding stage as follows: $\hat{i} = (i + K_1) \mod 2^{k\RSc}$ and $\hat{l} = (l + K_2) \mod 2^{k\RUc}$. Note that encryption of $i$ and $l$ can be partial because $R_{k1} \leq \RSc$ and $R_{k2} \leq \RUc$
}

\subsubsection{Channel Encoding Strategy}
\label{subsec:ch-enc}
Define $(i',j',l',p') = g(\hat{i},j,\hat{l},p)$ to be one-to-one and invertible mapping such that there exist sub-mappings $i' = g_1({\color{blue}\hat{i}})$, $(i',j') = g_2({\color{blue}\hat{i}},j)$, and $(l',p') = g_3({\color{blue}\hat{l}},p)$. Hence, rates should satisfy:
\begin{align}
\label{ineq:mapping-rates-1}
    \RSc &\leq \RAcc, \\
    \RSc + \RSp &\leq \RAcc + \RApp, \\
    \RUc + \RUp &\leq \RBcc + \RBpp.
\label{ineq:mapping-rates-3}
\end{align}
The channel encoding is as follows. Given indices from the source encoding {\color{blue} and encryption stages}, get $i'$, $j'$, $l'$, and $p'$ using mapping $g$. Randomly {\color{blue} and independently} select $\ws \in [1,...,2^{k\Rms}]$ and $\wu \in [1,...,2^{k\Rmu}]$ from the uniform distribution. {\color{blue} Select $\qsns,\quns,\wsns,\wuns$, pick $x^n$ and transmit it. For details on how $x^n$ is generated, see final step of the channel codebook generation in Appendix~\ref{appendix:codebook}}. See Fig.~\ref{fig:operational-separation} for the whole coding process illustration.
\ifCLASSOPTIONdraft
\begin{figure}
    \centering
    \includegraphics[width=0.5\linewidth, draft=false]{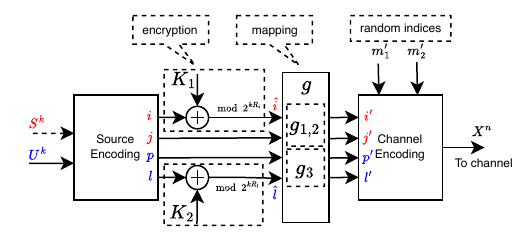}
    \caption{\color{blue} Relation between source and channel coding. The indices $i$ and $l$ are encrypted as $\hat{i} = (i + K_1 \mod 2^{k\RSc})$ and $\hat{l} = (l + K_2 \mod 2^{k\RUc})$, respectively}
    \label{fig:operational-separation}
\end{figure}
\else
\begin{figure}
    \centering
    \includegraphics[width=1\linewidth, draft=false]{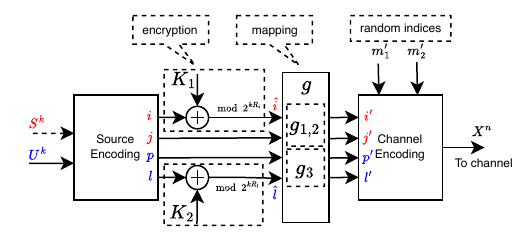}
    \caption{\color{blue} Relation between source and channel coding. The indices $i$ and $l$ are encrypted as $\hat{i} = (i + K_1 \mod 2^{k\RSc})$ and $\hat{l} = (l + K_2 \mod 2^{k\RUc})$, respectively}
    \label{fig:operational-separation}
\end{figure}
\fi
We assume that the mapping $g$, and the codebooks are revealed to both the legitimate receiver and the eavesdropper. {\color{blue} The secret key $K=(K_1,K_2)$ is only revealed to the legitimate receiver.}

\subsubsection{Decoding}
The receiver observes $y^n$ and sequentially searches for the unique codewords s.t.:
\begin{enumerate} 
    \item $(\qsnd, y^n) \in \T^n(\Acc, Y)$,
    \item $(\qund, y^n) \in \T^n(\Bcc, Y | \qsns)$,
    \item $(\wsnd, y^n) \in \T^n(\App, Y | \qsns)$,
    \item $(\wund, y^n) \in \T^n(\Bpp, Y | \qsns, \quns)$,
\end{enumerate}

The decoding will be successful {\color{blue} if only one true codeword is jointly typical with channel output for the legitimate receiver, which will hold} if the rates satisfy \cite[Packing Lemma 3.1]{elgamal2011book}:
\begin{align}
\label{ineq:rate-R'i}
    \RAcc &< (r + \epsilon) I(\Acc;Y) + \delta_{\color{blue} k},
\end{align}
\begin{align}
\label{ineq:rate-R'l}
    \RBcc &< (r + \epsilon) I(\Bcc;Y|\Acc) + \delta_{\color{blue} k}, \\
\label{ineq:rate-R'j+R'm1}
    \RApp + \Rms &< (r + \epsilon) I(\App;Y|\Acc) + \delta_{\color{blue} k}, \\
\label{ineq:rate-R'p+R'm2}
    \RBpp + \Rmu &< (r + \epsilon) I(\Bpp;Y|\Acc,\Bcc) + \delta_{\color{blue} k},
\end{align}
{\color{blue} where $\delta_{\color{blue} k} \to 0$ with $k \to \infty$.}

\subsubsection{Distortion Analysis}
In this subsection, we derive the region for expected distortion for the {\color{blue} semantic} part of the source.
The distortion analysis for the {\color{blue} observed} part follows the same steps and is omitted here for brevity. We bound multi-letter average distortion as follows:
{\color{blue} 
\ifCLASSOPTIONdraft
\begin{align}
    \EX d_s (S^k, \hat{S}^k) &= \EX d_s (S^k, \hat{f}_s(Y^n, {\color{blue} K})) \\
    &{=^{(a)}} P_{\Err} \EX \left\{ d_s(S^k, \hat{f}_s(Y^n, {\color{blue} K})) | \Err \right\} {+} P_{\bar{\Err}} \EX \left\{ d_s(S^k, \hat{f}_s(Y^n, {\color{blue} K})) | \bar{\Err} \right\} \nonumber \\
    &\leq P_{\Err} D_m + P_{\bar{\Err}} \EX \left\{ d_s( S^k, \hat{f}_s(Y^n, {\color{blue} K})) | \bar{\Err} \right\} \\
    &=^{(b)} P_{\Err} D_m + P_{\bar{\Err}} \EX \left\{ d_s( S^k, \tilde{S}(\Sc^k,\Sp^k)) | \bar{\Err} \right\} \\
    &\leq^{(c)} P_{\Err} D_m {+} P_{\bar{\Err}} (1 {+} \epsilon_1) \EX \left\{ d_s(S, \tilde{S}(\Sc,\Sp)) \right\},
\end{align}
\else
\begin{align}
    &\EX d_s (S^k, \hat{S}^k) = \EX d_s (S^k, \hat{f}_s(Y^n, {\color{blue} K})) \\
    &{=^{(a)}} P_{\Err} \EX \left\{ d_s(S^k, \hat{f}_s(Y^n, {\color{blue} K})) | \Err \right\} \nonumber \\ &\quad + P_{\bar{\Err}} \EX \left\{ d_s(S^k, \hat{f}_s(Y^n, {\color{blue} K})) | \bar{\Err} \right\}  \\
    &\leq P_{\Err} D_m + P_{\bar{\Err}} \EX \left\{ d_s( S^k, \hat{f}_s(Y^n, {\color{blue} K})) | \bar{\Err} \right\} \\
    &=^{(b)} P_{\Err} D_m + P_{\bar{\Err}} \EX \left\{ d_s( S^k, \tilde{S}(\Sc^k,\Sp^k)) | \bar{\Err} \right\} \\
    &\leq^{(c)} P_{\Err} D_m {+} P_{\bar{\Err}} (1 {+} \epsilon_1) \EX \left\{ d_s(S, \tilde{S}(\Sc,\Sp)) \right\},
\end{align}
\fi
}
where ${\color{blue} \hat{f}_s : \mathcal{Y}^n  \times \mathcal{K} \to \mathcal{S}^k}$ is the decoding function for observation (Definition~\ref{def:decoder}){\color{blue}, $\Err$ is an error event defined in Appendix~\ref{appendix:errors},} (a) due to the law of total expectation, $D_m = {\color{blue}\max_{s^k, \hat{s}^k} d_s(s^k,\hat{s}^k)}$ is the maximum distortion, (b) is because ${\color{blue} \hat{f}_s(Y^n, K) = \tilde{S}(\Sc^k,\Sp^k)}$ given that no error occurs $\bar{\Err}$, where ${\color{blue} \tilde{S}(.)}$ is the function that reconstructs observation ${\color{blue} S^k}$ from source encoder sequences $\Sc^k$ and ${\color{blue} \Sp^k}$, (c) {\color{blue} is the transition from multi-letter average to the single-letter one, and it is} due to typical average lemma {\color{blue} \cite{elgamal2011book} given that $({\color{blue} S^k},\Sc^k,\Sp^k) \in \T^k({\color{blue} S},\Sc,\Sp)$, which in Case 1 encoding holds due to Markov Lemma \cite{elgamal2011book} given that $S \to U \to (\Sc,\Sp)$.} 
With $P_{\Err} \to 0$ we have:
\begin{align}
    \EX d_s(S^k, \hat{S}^k) \leq (1 + \epsilon_1)\EX
    d_s(S, \tilde{S} (\Sc,\Sp))
\end{align}
Hence, if $D_s \leq \EX d_s(S, \tilde{S} (\Sc,\Sp)) - \epsilon$, then the distortion constraint for the {\color{blue} semantic part (\ref{eq:sem-distortion}}) is satisfied. Similarly, it can also be shown that if {\color{blue} $D_u \leq \EX d_u(U, \tilde{U} (\Sc,\Uc,\Up)) - \epsilon$}, then the distortion for the {\color{blue} observed} part ({\color{blue}\ref{eq:obs-distortion}}) is satisfied.

\subsubsection{Equivocation Analysis}
\label{subsubsec:eqv-analysis}
In this subsection, we analyse the equivocation at the eavesdropper for the proposed coding scheme. We start with the derivation for the semantic part. Firstly, we expand and lower bound equivocation for the semantic part in the following way. 
\ifCLASSOPTIONdraft
\begin{align}
    &H(S^k|Z^n) \geq^{(a)} H(S^k) - H(C) \ { \color{blue} + \ H(C_0|S^k) } - I(Z^n;S^k|C) + H({\App}^n|C_0) - H({\App}^n|C) \nonumber \\
    &\quad- H(Z^n|C_0) + H(Z^n|C, {\App}^n),
\label{eq:sem-eqv-terms}
\end{align}
\else
\begin{align}
    &H(S^k|Z^n) \geq^{(a)} H(S^k) - H(C) \ { \color{blue} + \ H(C_0|S^k) } \nonumber \\ &\quad- I(Z^n;S^k|C) + H({\App}^n|C_0) - H({\App}^n|C) \nonumber \\
    &\quad- H(Z^n|C_0) + H(Z^n|C, {\App}^n),
\label{eq:sem-eqv-terms}
\end{align}
\fi
where the proof of (a) can be found in Appendix~\ref{appendix:eqv-ineq}.

This expression is valid for arbitrary r.v. $C = (C_0,C_1)$. We set $C = (I',J')$, where $I'$ and $J'$ are r.v.s that represent random indices $i'$ and $j'$ in the channel codebook. Then, the first {\color{blue} three} terms of (\ref{eq:sem-eqv-terms}) reflect loss in the source encoding {\color{blue} and the effect of the encryption} of the semantic part and can be lower bounded as follows:
\ifCLASSOPTIONdraft
\begin{align}
    &H(S^k) - H(I',J') \ { \color{blue} + H(I'|S^k) } =^{(a)} H(S^k) - H({\color{blue} \hat{I}},J) \ {\color{blue} + \ H(\hat{I}|S^k)} \\ &\quad \geq^{(b)} k H(S) - k\RSc - k\RSp \ {\color{blue} + \ kR_{k1} },
\end{align}
\else
\begin{align}
    &H(S^k) - H(I',J') + H(I'|S^k) \overset{(a)}{=} H(S^k) - H(\hat{I},J) \nonumber \\ &\quad+ H(\hat{I}|S^k) \overset{(b)}{\geq} k H(S) - k\RSc - k\RSp +  kR_{k1},
\end{align}
\fi
where $\hat{I}$ and $J$ are r.v.s representing random indices {\color{blue}$\hat{i}$ (encrypted)} and $j$ in the source codebook, (a) is due to the existence of mappings {\color{blue} $g_1^{-1}$ and } $g_2^{-1}$, defined above in \ref{subsec:ch-enc}, (b) is because $S^k$ is i.i.d, $ H({\color{blue} \hat{I}},J) \leq H({ \color{blue} \hat{I} }) + H(J) \leq k(\RSc + \RSp)$, {\color{blue} and $H(\hat{I}|S^k) \geq H(I + K_1 \text{ mod } 2^{k\RSc}|S^k, U^k) = H(K_1|S^k,U^k) = H(K_1) = kR_{k1}$ because $I$ is a function of $(S^k,U^k)$ and $K_1$ is uniform and independent of $(S^k,U^k)$. Additionally, $I(Z^n;S^k|I',J') \leq k I(S;\B|\A) + \epsilon$ and the proof of this bound is in Appendix~\ref{appendix:mi-term}.} 

The remaining terms of (\ref{eq:sem-eqv-terms}) depend on the channel codebook and have the following bounds \cite{liang2009}:
\begin{enumerate}
    \item $H({\App}^n|I') \geq^{(a)}  k\RApp + k\Rms - 1 - \epsilon$,
    \item $H({\App}^n|I',J') \leq k\Rms$,
    \item $H(Z^n|I') \leq^{(b)}  n H(Z|\Acc) + \epsilon$,
    \item $H(Z^n|I',J',{\App}^n) {=^{(c)}}  n H(Z|\App) {\geq} nH(Z|\Acc,\App)$
\end{enumerate}
where (a) due to \cite[Lemma 2.5]{liang2009}, (b) for the same reasons as in \cite[eq. (2.50)-(2.54)]{liang2009}, (c) similarly to \cite[eq. (2.43)-(2.46)]{liang2009}.

Returning to (\ref{eq:sem-eqv-terms}) we obtain:
\ifCLASSOPTIONdraft
\begin{align}
    &\frac{1}{k} H(S^k|Z^n) \geq  H(S) {\color{blue} \ - \ I(S;\B|\A)} - \RSc - \RSp \ {\color{blue} +  R_{k1}} + \RApp + \tfrac{n}{k}H(Z|\Acc,\App) - \tfrac{n}{k}H(Z|\Acc) \nonumber \\
    &\quad - \tfrac{1+\epsilon}{k} 
    \geq^{(a)} H(S) {\color{blue} \ - \ I(S;\B|\A)} - \RSc - \RSp \ {\color{blue} +  R_{k1}} + \RApp - (r + \epsilon)I(\App;Z|\Acc) - {\color{blue} \delta_k},
\end{align}
\else
\begin{align}
    &\frac{1}{k} H(S^k|Z^n) \geq  H(S) {\color{blue} \ - \ I(S;\B|\A)} - \RSc - \RSp \ {\color{blue} +  R_{k1}} \nonumber \\ 
    &\quad + \RApp + \tfrac{n}{k}H(Z|\Acc,\App) - \tfrac{n}{k}H(Z|\Acc) - \tfrac{1+\epsilon}{k}  \\
    &\geq^{(a)} H(S) {\color{blue} \ - \ I(S;\B|\A)} - \RSc - \RSp \ {\color{blue} + R_{k1}} + \RApp \nonumber \\
    &\quad - (r + \epsilon)I(\App;Z|\Acc) - {\color{blue} \delta_k},
\end{align}
\fi
where (a) follows from (\ref{ineq:rate-cond}); ${\color{blue} \delta_k} \to 0$ with $k \to \infty$. This lower bound implies that equivocation for the semantic part is achievable for the proposed coding scheme if
\ifCLASSOPTIONdraft
\begin{align}
    \Delta_s &\leq H(S) {\color{blue} \ - \ I(S;\B|\A)} - \RSc - \RSp \ {\color{blue} +  R_{k1}} + \RApp - (r + \epsilon)I(\App;Z|\Acc) - {\color{blue} \delta_k}.
    \label{ineq:proof-sem-eqv}
\end{align}
\else
\begin{align}
    \Delta_s &\leq H(S) {\color{blue} \ - \ I(S;\B|\A)} - \RSc - \RSp \ {\color{blue} +  R_{k1}} + \RApp \nonumber 
    \\ &\quad - (r + \epsilon)I(\App;Z|\Acc) - {\color{blue} \delta_k}.
    \label{ineq:proof-sem-eqv}
\end{align}
\fi

Following similar steps, for the equivocation of the observation, we have:
\ifCLASSOPTIONdraft
\begin{align}
    &H(U^k|Z^n) \geq H(U^k) - H(C) \ { \color{blue} + \ H(C_0|U^k)} - I(Z^n;U^k|C) + H({\Bpp}^n|C_0) - H({\Bpp}^n|C) \nonumber \\
    &\quad - H(Z^n|C_0) + H(Z^n|C,{\Bpp}^n),
    \label{eq:obs-eqv-terms}
\end{align}
\else
\begin{align}
    &H(U^k|Z^n) \geq H(U^k) - H(C) \ { \color{blue} + \ H(C_0|U^k)} \nonumber \\
    &\quad - I(Z^n;U^k|C) + H({\Bpp}^n|C_0) - H({\Bpp}^n|C) \nonumber \\
    &\quad - H(Z^n|C_0) + H(Z^n|C,{\Bpp}^n),
    \label{eq:obs-eqv-terms}
\end{align}
\fi
where $C_0 = (I',L')$ and $C_1 = P'$, with $L'$ and $P'$ representing the random indices $l'$ and $p'$ from the channel codebook. The first {\color{blue} three} terms on the right-hand side of (\ref{eq:obs-eqv-terms}) are related to the source encoding and encryption and have the following lower bound:
\ifCLASSOPTIONdraft
\begin{align}
    &H(U^k) - H(I',L',P') \ { \color{blue} + \ H(I',L'|U^k)} =^{(a)} H(U^k) - H({ \color{blue}\hat{I},\hat{L}},P) \ { \color{blue} + \ H(\hat{I},\hat{L}|U^k)} \\
    &\geq^{(b)} kH(U) - k\RSc - k\RUc - k\RUp \ {\color{blue} + \ kR_{k1} + kR_{k2}},
\end{align}
\else
\begin{align}
    &H(U^k) - H(I',L',P') \ { \color{blue} + \ H(I',L'|U^k)} \\ &=^{(a)} H(U^k) - H({ \color{blue}\hat{I},\hat{L}},P) \ { \color{blue} + \ H(\hat{I},\hat{L}|U^k)} \\
    &\geq^{(b)} kH(U) - k\RSc - k\RUc - k\RUp \ {\color{teal} + \ kR_{k}},
\end{align}
\fi
where $\hat{L}$ and $P$ represent the random source codebook indices $\hat{l}$ {\color{blue} (encrypted)} and $p$, respectively, (a) is due to $g^{-1}_1$ and $g^{-1}_3$ mappings defied in \ref{subsec:ch-enc}, (b) holds because $U^k$ is i.i.d., $H(C) = H({\color{blue} \hat{I},\hat{L}},P) \leq k\RSc + k\RUc + k\RUp$, and {\color{blue} $H(\hat{I},\hat{L}|U^k) \geq H(I + K_1 \text{ mod } 2^{k\RSc},L + K_2 \text{ mod } 2^{k\RUc}|S^k,U^k) = kR_{k1} + kR_{k2}$.} {\color{blue} Also, we have $I(Z^n;U^k|I',L',P') \leq^{(a)} k I(U;\Sp|\Sc,\B) + \epsilon$  proof of which is outlined in Appendix~\ref{appendix:mi-term}.}

Next, we have the following bounds for each of the remaining terms of (\ref{eq:obs-eqv-terms}).
\begin{enumerate}
    \item $H({\Bpp}^n|I',L') \geq^{(a)}  k\RBpp + k\Rmu - 1 - \epsilon$,
    \item $H({\Bpp}^n|I',L',P') \leq k\Rmu$,
    \item $H(Z^n|I',L') \leq^{(b)}  n H(Z|\Acc,\Bcc) + \epsilon$,
    \item $H(Z^n|I',L',P',{\Bpp}^n) =^{(c)}  n H(Z|\Bpp) \geq n H(Z|\Acc,\Bcc,\Bpp)$,
\end{enumerate}
where (a) is due to \cite[Lemma 2.5]{liang2009}, (b) for the same reasons as in \cite[eq. (2.50)-(2.54)]{liang2009}, (c) similarly to \cite[eq. (2.43)-(2.46)]{liang2009}.
Hence, for (\ref{eq:obs-eqv-terms}) we have:
\ifCLASSOPTIONdraft
\begin{align}
    &\frac{1}{k} H(U^k|Z^n) \geq  H(U) \ {\color{blue} - \ I(U;\Sp|\Sc,\B)} - \RUc - \RUp \ {\color{blue} + \ R_{k}} + \RBpp + \tfrac{n}{k}H(Z|\Acc,\Bcc,\Bpp) \nonumber \\
    &\quad - \tfrac{n}{k}H(Z|\Acc,\Bcc) - \tfrac{1+\epsilon}{k}
    \geq^{(a)} H(U) \ {\color{blue} - \ I(U;\Sp|\Sc,\B)} - \RUc - \RUp \ {\color{blue} + \ R_{k} } + \RBpp \nonumber \\ 
    &\quad - (r + \epsilon)I(\Bpp;Z|\Acc,\Bcc) - {\color{blue} \delta_k},
\end{align}
\else
\begin{align}
    &\frac{1}{k} H(U^k|Z^n) \geq  H(U) \ {\color{blue} - \ I(U;\Sp|\Sc,\B)} - \RUc - \RUp \nonumber \\
    &\quad + R_{k} + \RBpp + \tfrac{n}{k}H(Z|\Acc,\Bcc,\Bpp) - \tfrac{n}{k}H(Z|\Acc,\Bcc) \nonumber \\
    &\quad- \tfrac{1+\epsilon}{k}
    \geq^{(a)} H(U) \ {\color{blue} - \ I(U;\Sp|\Sc,\B)} - \RUc - \RUp \nonumber \\
    &\quad + R_{k} + \RBpp - (r + \epsilon)I(\Bpp;Z|\Acc,\Bcc) - {\color{blue} \delta_k},
\end{align}
\fi
where (a) due to (\ref{ineq:rate-cond}); ${\color{blue} \delta_k} \to 0$ with $k \to \infty$. This lower bound implies that the equivocation $\Delta_u$ for the observed part is achievable in the proposed coding scheme if,
\ifCLASSOPTIONdraft
\begin{align}
    &\Delta_u \leq H(U) \ {\color{blue} - \ I(U;\Sp|\Sc,\B)} - \RUc - \RUp \ {\color{blue} + \ R_{k} } + \RBpp - (r + \epsilon)I(\Bpp;Z|\ABcc) - {\color{blue} \delta_k}.
    \label{ineq:proof-obs-eqv}
\end{align}
\else
\begin{align}
    &\Delta_u \leq H(U) \ {\color{blue} - \ I(U;\Sp|\Sc,\B)} - \RUc - \RUp \nonumber \\
    &\quad + R_{k} + \RBpp - (r + \epsilon)I(\Bpp;Z|\ABcc) - {\color{blue} \delta_k}.
    \label{ineq:proof-obs-eqv}
\end{align}
\fi
Similarly to the derivations for the semantic and observed parts, we obtain the bound for the joint equivocation:
\ifCLASSOPTIONdraft
\begin{align}
    &H(S^k,U^k|Z^n) \geq H(S^k,U^k) - H(C) \ { \color{blue} + \ H(C_0|S^k,U^k) } - I(Z^n;S^k,U^k|C) + H({\App}^n,{\Bpp}^n|C_0)  \nonumber \\
    &\quad- H({\App}^n,{\Bpp}^n|C) - H(Z^n|C_0) + H(Z^n|C,{\App}^n,{\Bpp}^n),
    \label{eq:joint-eqv-terms}
\end{align}
\else
\begin{align}
    &H(S^k,U^k|Z^n) {\geq} H(S^k,U^k) - H(C) \ { \color{blue} + \ H(C_0|S^k,U^k) } \nonumber \\
    &\quad - I(Z^n;S^k,U^k|C) + H({\App}^n,{\Bpp}^n|C_0) \nonumber \\
    &\quad- H({\App}^n,{\Bpp}^n|C) - H(Z^n|C_0) \nonumber \\
    &\quad + H(Z^n|C,{\App}^n,{\Bpp}^n),
    \label{eq:joint-eqv-terms}
\end{align}
\fi
where {\color{blue} we set $C_0 = (I',L')$, and $C_1 = (J',P')$.} The first {\color{blue} three} terms of the right hand side of (\ref{eq:joint-eqv-terms}) {\color{blue} can be lower bounded as}
\ifCLASSOPTIONdraft
\begin{align}
    &H(S^k,U^k) - H(C) \ { \color{blue} + \ H(C_0|S^k,U^k) } \geq^{(a)} kH(S,U) - k\RSc - k\RSp - k\RUc - k\RUp + {\color{blue} kR_{k},}
\end{align}
\else
\begin{align}
    &H(S^k,U^k) - H(C) \ { \color{blue} + \ H(C_0|S^k,U^k) } \\
    &\geq^{(a)} kH(S{,}U) {-} k\RSc {-} k\RSp {-} k\RUc {-} k\RUp {+} {\color{blue} kR_{k},}
\end{align}
\fi
where (a) is due to $(S^k,U^k)$ are i.i.d and $g^{-1}$ mapping {\color{blue} and secret key rates similarly to previous derivation for the observation equivocation. The forth term is zero due to $(S^k,U^k) \to (I',J',L',P') \to Z^n$ Markov chain.}
The remaining channel encoder terms in (\ref{eq:joint-eqv-terms}) are
\begin{enumerate}
    \item $H({\App}^n,{\Bpp}^n|I',L') \geq k\RApp + k\RBpp + k\Rms + k\Rmu - 1 - \epsilon$,
    \item $H({\App}^n,{\Bpp}^n|I',J',L',P') \leq k\Rms + k\Rmu$,
    \item $H(Z^n|I',L') \leq n H(Z|\ABcc) + \epsilon$,
    \item $H(Z^n|I',J',L',P',{\App}^n,{\Bpp}^n) = n H(Z|\ABpp) \geq n H(Z|\ABcc,\ABpp)$.
\end{enumerate}
Hence, we have
\ifCLASSOPTIONdraft
\begin{align}
    &\frac{1}{k} H(S^k,U^k|Z^n) \geq  H(S,U) - \RSc - \RSp - \RUc - \RUp {\color{blue} \ + \ R_{k}} + \RApp + \RBpp \nonumber \\
    &\quad + \tfrac{n}{k}H(Z|\ABcc,\ABpp) - \tfrac{n}{k}H(Z|\ABcc) - \tfrac{1+2\epsilon}{k}
    \geq^{(a)} H(S,U) - \RSc - \RSp - \RUc - \RUp {\color{blue} \ + \ R_{k}} \nonumber \\
    &\quad + \RApp + \RBpp - (r + \epsilon)I(\ABpp;Z|\ABcc) - \epsilon_k,
\end{align}
\else
\begin{align}
    &\frac{1}{k} H(S^k,U^k|Z^n) \geq  H(S,U) - \RSc - \RSp - \RUc - \RUp \nonumber \\
    &\quad {\color{blue} \ + \ R_{k}} + \RApp + \RBpp + \tfrac{n}{k}H(Z|\ABcc,\ABpp) - \tfrac{n}{k}H(Z|\ABcc) \nonumber \\
    &\quad - \tfrac{1+2\epsilon}{k}
     \geq^{(a)} H(S,U) - \RSc - \RSp - \RUc - \RUp {\color{blue} \ + \ R_{k}} \nonumber \\
    &\quad + \RApp + \RBpp - (r + \epsilon)I(\ABpp;Z|\ABcc) - \epsilon_k,
\end{align}
\fi
where (a) follows from (\ref{ineq:rate-cond}); $\epsilon_k \to 0$ with $k \to \infty$. Hence, the joint equivocation is achievable if:
\ifCLASSOPTIONdraft
\begin{align}
    &\Delta_{su} \leq H(S,U) - \RSc - \RSp - \RUc - \RUp {\color{blue} \ + \ R_{k}} + \RApp + \RBpp - (r + \epsilon)I(\ABpp;Z|\ABcc) - \epsilon_k.
    \label{ineq:proof-joint-eqv}
\end{align}
\else
\begin{align}
    &\Delta_{su} \leq H(S,U) - \RSc - \RSp - \RUc - \RUp {\color{blue} \ + \ R_{k}} \nonumber \\
    &\quad + \RApp + \RBpp - (r + \epsilon)I(\ABpp;Z|\ABcc) - \epsilon_k.
    \label{ineq:proof-joint-eqv}
\end{align}
\fi

\subsubsection{Summary of Coding Scheme Conditions}
The conditions for the proposed achievability coding scheme are presented in (\ref{ineq:rate-Ri-Rj-entropy})-(\ref{ineq:rate-Rl-Rp-entropy}), (\ref{ineq:rate-Ri})-(\ref{ineq:rate-Rp}),
(\ref{ineq:ch-cb-Ws-rate})-(\ref{ineq:ch-cb-Qs-rate}), (\ref{ineq:mapping-rates-1})-(\ref{ineq:mapping-rates-3}),
(\ref{ineq:rate-R'i})-(\ref{ineq:rate-R'p+R'm2}),
(\ref{ineq:proof-sem-eqv}),
(\ref{ineq:proof-obs-eqv}),
(\ref{ineq:proof-joint-eqv}){\color{blue}, as well as in the encryption process description}. The system of inequalities formed by {\color{teal} the scheme} conditions can be reduced using Fourier-Motzkin elimination \cite{elgamal2011book} given $\epsilon \to 0$ and $k \to \infty$,
and with straightforward simplifications, we {\color{teal} can} obtain the form presented in Theorem~\ref{theorem:direct}. This completes the proof of achievability.
\section{Special Cases: Gaussian and Binary Models} \label{sec:gauss-and-binary}
In this section, we analyse the fundamental limits for the transmission of bivariate Gaussian and Bernoulli sources over Gaussian and binary wiretap channels, providing a {\color{blue} closed-from} characterisation of the rate-distortion-equivocation trade-off.

\subsection{Gaussian System Model}
\label{sub:gauss-model}

Consider a bivariate Gaussian source $(S,U) \sim \mathcal{N}(\mathbf{0},\mathbf{K})$ with a covariance matrix:
\begin{equation}
    \mathbf{K} =
    \begin{pmatrix}
        P_s & P_{su} \\
        P_{su} & P_u
    \end{pmatrix},
\end{equation}
where $P_s, P_u > 0$ are source component powers and $P_{su}$ is their covariance. We set quadratic distortion measures for both parts of the source: $d_s(x,\hat{x}) = d_u(x,\hat{x}) =(x-\hat{x})^2$. We model the Gaussian wiretap channel as:
\begin{equation}
\begin{cases}
Y = X + N_1, & N_1 \sim \mathcal{N}(0,P_{N_1}) \\
Z = Y + N_2, & N_2 \sim \mathcal{N}(0,P_{N_2})
\end{cases}
\end{equation}
with channel input power constraint $\EX[X^2] \leq P$. The noise power of the eavesdropper channel is $P_N = P_{N_1} + P_{N_2}$.

The definition of the achievable tuple for the continuous Gaussian case differs from the general Definition~\ref{def:achievable}. Specifically, we use the differential entropy $h(.)$ instead of the discrete $H(.)$ and introduce the channel input power constraint. 

\begin{definition}[Achievable Tuple] \label{def:gauss-achievable-tuple}
A tuple \( (r, P, R_k, D_s, D_u, \Delta_s, \Delta_u, \Delta_{su}) \in {\color{blue} \mathbb{R}_+^8} \) is achievable for the Gaussian system model if for any \( \epsilon > 0 \), there exists a \( (k,n) \)-code satisfying:
\ifCLASSOPTIONdraft
\begin{align}
    \label{ineq:gauss-rate-constraint}
    \frac{n}{k} &\leq r + \epsilon
    \hspace{7mm} {\text{\rm(rate constraint)}} \\
    \EX[X^2] &\leq P + \epsilon \hspace{7mm} {\text{\rm(power constraint)}} \\
    \frac{1}{k}\log|\mathcal{K}| &\leq R_k + \epsilon
    \hspace{5mm} {\text{\rm(key rate constraint)}} \\
    \frac{1}{k} \EX \sum_{i=1}^{k} (S_i - \hat{S}_i)^2 &\leq D_s + \epsilon
    \hspace{5mm} {\text{\rm(semantic distortion)}}  \\
    \frac{1}{k} \EX \sum_{i=1}^{k} (U_i - \hat{U}_i)^2 &\leq D_u + \epsilon
    \hspace{5mm} 
    {\text{\rm(observed part distortion)}} 
    \end{align}
    \begin{align}
    \frac{1}{k}h(S^k|Z^n) &\geq \Delta_s - \epsilon
    \hspace{5mm} {\text{\rm(semantic part secrecy)}}
    \label{ineq:guass-sem-eqv-def} \\
    \frac{1}{k}h(U^k|Z^n) &\geq \Delta_u - \epsilon
    \hspace{5mm} {\text{\rm(observed part secrecy)}} \\
    \frac{1}{k}h(S^k,U^k|Z^n) &\geq \Delta_{su} - \epsilon
    \hspace{4mm} {\text{\rm(joint secrecy)}}
\end{align}
\else
\begin{align}
    \label{ineq:gauss-rate-constraint}
    &\frac{n}{k} \leq r + \epsilon
    \hspace{1mm} {\text{\rm(rate constraint)}} \\
    &\EX[X^2] \leq P + \epsilon \hspace{1mm} {\text{\rm(power constraint)}} \\
    &\frac{1}{k}\log|\mathcal{K}| \leq R_k + \epsilon
    \hspace{1mm} {\text{\rm(key rate constraint)}} \\
    &\frac{1}{k} \EX \sum_{i=1}^{k} (S_i - \hat{S}_i)^2 \leq D_s + \epsilon
    \hspace{1mm} {\text{\rm(sem. distortion)}}  \\
    &\frac{1}{k} \EX \sum_{i=1}^{k} (U_i - \hat{U}_i)^2 \leq D_u + \epsilon
    \hspace{1mm} 
    {\text{\rm(obs. part dist.)}} \\
    &\frac{1}{k}h(S^k|Z^n) \geq \Delta_s - \epsilon
    \hspace{1mm} {\text{\rm(sem. part secrecy)}}
    \label{ineq:guass-sem-eqv-def} \\
    &\frac{1}{k}h(U^k|Z^n) \geq \Delta_u - \epsilon
    \hspace{1mm} {\text{\rm(obs. part secrecy)}} \\
    &\frac{1}{k}h(S^k,U^k|Z^n) \geq \Delta_{su} - \epsilon
    \hspace{1mm} {\text{\rm(joint secrecy)}}
\end{align}
\fi
\end{definition}
\begin{remark}
    The secret key $K$ remains to be discrete r.v. with finite alphabet.
\end{remark}

\label{sub:gauss-outer}
 The following corollary follows from Theorem~\ref{theorem:converse}.
 
\begin{corollary}[Gaussian Converse Bound] \label{corollary:gauss-outer}
For the Gaussian system described in Subsection~\ref{sub:gauss-model}, any achievable tuple $(r, P, R_k, D_s, D_u, \Delta_s, \Delta_u, \Delta_{su})$ must satisfy the following conditions:
\ifCLASSOPTIONdraft
\begin{align}
    &R(D_s,D_u) \leq \frac{r}{2} \log \left( 1 + \frac{P}{P_{N_1}} \right), \\
    \label{ineq:gaus-converse-sem-eqv}
    \Delta_s &\leq R_k + \frac{r}{2} \left[ \log \left( 1 + \frac{\beta_1 P}{P_{N_1}} \right) - \log \left( 1 + \frac{\beta_1 P}{P_{N}} \right) \right] + \frac{1}{2} \log 2 \pi e P_s - R_s(D_s), \\
    \label{ineq:gaus-converse-src-eqv}
    \Delta_u &\leq R_k + \frac{r}{2} \left[ \log \left( 1 + \frac{\beta_2 P}{P_{N_1}} \right) - \log \left( 1 + \frac{\beta_2 P}{P_{N}} \right) \right] + \frac{1}{2} \log 2 \pi e P_u - R_u(D_u), \\
    \label{ineq:gaus-converse-joint-eqv}
    \Delta_{su} &\leq R_k + \frac{r}{2} \left[ \log \left( 1 + \frac{P}{P_{N_1}} \right) - \log \left( 1 + \frac{P}{P_{N}} \right) \right] + \frac{1}{2} \log (2 \pi e)^2 |\mathbf{K}| - R(D_s,D_u),
\end{align}
\else
\begin{align}
    &R(D_s,D_u) \leq \frac{r}{2} \log \left( 1 + \frac{P}{P_{N_1}} \right), 
    \\
    \label{ineq:gaus-converse-sem-eqv}
    \Delta_s &\leq R_k + \frac{r}{2} \left[ \log \left( 1 + \frac{\beta_1 P}{P_{N_1}} \right) - \log \left( 1 + \frac{\beta_1 P}{P_{N}} \right) \right] \nonumber \\
    &\quad + \frac{1}{2} \log 2 \pi e P_s - R_s(D_s), \\
    \label{ineq:gaus-converse-src-eqv}
    \Delta_u &\leq R_k + \frac{r}{2} \left[ \log \left( 1 + \frac{\beta_2 P}{P_{N_1}} \right) - \log \left( 1 + \frac{\beta_2 P}{P_{N}} \right) \right] \nonumber \\
    &\quad + \frac{1}{2} \log 2 \pi e P_u - R_u(D_u),
\end{align}
\begin{align}
    \label{ineq:gaus-converse-joint-eqv}
    \Delta_{su} &\leq R_k + \frac{r}{2} \left[ \log \left( 1 + \frac{P}{P_{N_1}} \right) - \log \left( 1 + \frac{P}{P_{N}} \right) \right] \nonumber \\
    &\quad + \frac{1}{2} \log (2 \pi e)^2 |\mathbf{K}| - R(D_s,D_u),
\end{align}
\fi
for some $\beta_1, \beta_2 \in [0,1]$, in Case 1 encoding $\beta_2 = 1$.
\end{corollary}
\begin{remark}
    There exists a threshold $\beta^*_1$ and $\beta^*_2$, such that for any $1 \geq  \beta_1 \geq \beta^*_1$ and $1 \geq \beta_2 \geq \beta^*_2$ converse bound is valid. Hence, one can always set $\beta_1 = \beta_2 = 1$ and obtain a valid converse bound.
    Also note that if we denote $C_s(\beta) =  \log \left( 1 + \frac{\beta P}{P_{N_1}} \right) - \log \left( 1 + \frac{\beta P}{P_{N}} \right)$ and define $C^*_s \doteq C_s(1)$, then $0 \leq C_s(\beta_1) + C_s(\beta_2) \leq 2 C^*_s$ for any $\beta_1, \beta_2 \in [0,1]$. 
\end{remark}

\subsubsection{Proof of Corollary \ref{corollary:gauss-outer}}

To prove the converse bound for the continuous Gaussian case, we rely on the proof of the general converse for the discrete r.v.s presented in Subsection~\ref{subsec:converse-proof}. Inequality (\ref{ineq:rd-region-proof-line-2}) and supporting ones (\ref{ineq:sem-rdf})-(\ref{ineq:ch-ineqs}) continue to hold for continuous r.v.s, so the proof for the rate-distortion region is the same as in Subsection~\ref{subsec:converse-proof}.

The proof of the equivocation region in the continuous case follows the same steps as for Theorem~\ref{theorem:converse} but it requires additional care with the entropy terms. Here we present a proof for the semantic equivocation region, and the region for observation and joint equivocation follows the same steps and is omitted.

Derivations (\ref{eq:sem-eqv-first-line}) remains to hold for Gaussian r.v.s because $K$ is the discrete r.v. We can continue from (\ref{eq:sem-eqv-first-line}) as follows.
\ifCLASSOPTIONdraft
\begin{align}
    H(K|Y^n) - H(K|Y^n,\hat{S}^k) &= I(K;\hat{S}^k|Y^n) = h(\hat{S}^k|Y^n) - h(\hat{S}^k|Y^n,K) = h(\hat{S}^k|Y^n) \\ &\geq^{(a)} h(\hat{S}^k|Y^n) - (h(S^k|Z^n) - k(\Delta_S - \epsilon)) \\ &\geq^{(b)} n I(\App;Z|\ABcc) - n I(\App;Y|\ABcc) - h(S^k) + I(S^k;\hat{S}^k) {+} k(\Delta_S {-} \epsilon),
\end{align}
\else
\begin{align}
    &H(K|Y^n) - H(K|Y^n,\hat{S}^k) \\
    &= I(K;\hat{S}^k|Y^n) = h(\hat{S}^k|Y^n) - h(\hat{S}^k|Y^n,K) \\ 
    &= h(\hat{S}^k|Y^n) \\ &\geq^{(a)} h(\hat{S}^k|Y^n) - (h(S^k|Z^n) - k(\Delta_S - \epsilon)) \\ &\geq^{(b)} n I(\App;Z|\ABcc) - n I(\App;Y|\ABcc) - h(S^k) \nonumber \\ &\quad + I(S^k;\hat{S}^k) {+} k(\Delta_S {-} \epsilon),
\end{align}
\fi
where (a) is due to (\ref{ineq:guass-sem-eqv-def}) and (b) is due to \cite[Lemma 17.12]{csiszar2011book}.
Using the result from the proof of \cite[Theorem 5.1]{bloch2011}, we bound the secrecy capacity term as follows:
\ifCLASSOPTIONdraft
\begin{align}
    n I(\App;Z) - n I(\App;Y) \geq \frac{n}{2} \log \left( 1 + \frac{\beta_1 P}{P_{N}} \right) - \frac{n}{2} \log \left( 1 + \frac{\beta_1 P}{P_{N_1}} \right),
\end{align}
\else
\begin{align}
    &n I(\App;Z) - n I(\App;Y) \\
    &\geq \frac{n}{2} \log \left( 1 + \tfrac{\beta_1 P}{P_{N}} \right) - \frac{n}{2} \log \left( 1 + \tfrac{\beta_1 P}{P_{N_1}} \right),
\end{align}
\fi
for some $\beta_1 \in [0, 1]$.
Also taking into account (\ref{ineq:sem-rdf}), (\ref{ineq:gauss-rate-constraint}), and the fact that $S^k$ is i.i.d we obtain:
\ifCLASSOPTIONdraft
\begin{align}
    (R_k + \epsilon) \geq \frac{1}{2} (r + \epsilon) \left[ \log \left( 1 + \frac{\beta_1 P}{P_{N}} \right) - \log \left( 1 + \frac{\beta_1 P}{P_{N_1}} \right) \right] - h(S) + R_s(D_s) + (\Delta_s - \epsilon)
\end{align}
\else
\begin{align}
    &(R_k + \epsilon) \geq \frac{1}{2} (r + \epsilon) \left[ \log \left( 1 {+} \tfrac{\beta_1 P}{P_{N}} \right) - \log \left( 1 {+} \tfrac{\beta_1 P}{P_{N_1}} \right) \right] \nonumber \\
    \quad & - h(S) + R_s(D_s) + (\Delta_s - \epsilon)
\end{align}
\fi
So we have the following bound for semantic equivocation, given $\epsilon \to 0$:
\ifCLASSOPTIONdraft
\begin{align}
    \Delta_s \leq R_k + \frac{r}{2} \left[ \log \left( 1 + \frac{\beta_1 P}{P_{N_1}} \right) - \log \left( 1 + \frac{\beta_1 P}{P_{N}} \right) \right] + \frac{1}{2} \log 2 \pi e P_s - R_s(D_s)
\end{align}
\else
\begin{align}
    &\Delta_s \leq R_k + \frac{r}{2} \left[ \log \left( 1 + \frac{\beta_1 P}{P_{N_1}} \right) - \log \left( 1 + \frac{\beta_1 P}{P_{N}} \right) \right] \nonumber \\
    &\quad + \frac{1}{2} \log 2 \pi e P_s - R_s(D_s)
\end{align}
\fi
Following the same steps, we can also obtain the bound for observation equivocation (\ref{ineq:gaus-converse-src-eqv}) and joint equivocation (\ref{ineq:gaus-converse-joint-eqv}). In the latter case, we set $\beta = 1$ in the secrecy capacity term.

\subsubsection{Rate-distortion functions} The converse bound in Corollary~\ref{corollary:gauss-outer} is valid for Case 1 and Case 2 of the encoder, however, explicit equations for RDFs $R_u(D_u)$, $R_s(D_s)$, and $R(D_s,D_u)$ depend on the encoder case. Specifically,
Lemma~\ref{lemma:marginal-gauss-rdfs} presents marginal RDFs (with a single distortion constraint) for the Case 1 and Case 2 encoder, while Lemma~\ref{lemma:joint-gaussian-rdfs} contains closed-form solutions for the joint RDF for Case 1 and Case 2 encoder.

\begin{lemma}[Marginal Gaussian RDFs] \label{lemma:marginal-gauss-rdfs}
Depending on the encoder input case as described in Section~\ref{sec:problem}, marginal RDFs when source is Gaussian have the following closed-form solutions: \\
\textbullet \ Case 1 and Case 2 RDF for the observation \cite{elgamal2011book}:
\begin{equation}
R_u(D_u) = \frac{1}{2}\log^+\frac{P_u}{D_u}.
\end{equation}
\textbullet \ Case 1 RDF for the semantic part \cite[Proposition 3]{liu2021}: 
\begin{equation}
R_{s,i}(D_s) = \frac{1}{2}\log^+\frac{\rho^2 P_s}{D_s - P_s(1 - \rho^2)},
\end{equation}
where $D_s > (1-\rho^2)P_s$. \\
\textbullet \ Case 2 RDF for the semantic part \cite{elgamal2011book}: 
\begin{equation}
R_{s,d}(D_s) = \frac{1}{2} \log^+ \frac{P_s}{D_s}.
\end{equation}
\end{lemma}

\begin{lemma}[Joint Gaussian RDFs] \label{lemma:joint-gaussian-rdfs}
Depending on the encoder input case, RDFs with separate distortions for each component of the Gaussian semantic source have the following form: \\
\textbullet \ Case 1 RDF \cite[Proposition 3]{liu2021}: 
\begin{equation}
R_i(D_s,D_u) = \max\left\{R_u(D_u), R_{s,i}(D_s)\right\},
\end{equation}
where $R_u(D_u)$ and $R_{s,i}(D_s)$ are given in Lemma~\ref{lemma:marginal-gauss-rdfs}. \\
\textbullet \ Case 2 RDF \cite[Theorem 6]{xiao2005}:

Let $\hat{D}_s = P_s - D_s$, and $\hat{D}_u = P_u - D_u$ then:
\begin{align}
R_d(D_s, D_u) &= \frac{1}{2} \log \frac{P_s}{D_s}, &\text{if } \rho^2 > \dfrac{\hat{D}_u P_s}{\hat{D}_s P_u}, \\
R_d(D_s, D_u) &= \frac{1}{2} \log \frac{|\mathbf{K}|}{D_s D_u}, &\text{if } \rho^2 < \dfrac{\hat{D}_s \hat{D}_u}{P_s P_u},
\end{align}
Otherwise,
\begin{equation}
R_d(D_s, D_u) = \frac{1}{2} \log \frac{|\mathbf{K}|}{D_s D_u - \left( \rho \sqrt{P_s P_u} - \sqrt{\hat{D}_s \hat{D}_u} \right)^2}.
\label{eq:case3}
\end{equation}
\end{lemma}

\subsection{Numerical Evaluations for the Gaussian System Model}
\label{sub:gauss-numerics}
\subsubsection{Converse Bound}
Fig.~\ref{subfig:gauss-outer-3D} presents a 3D plot of the converse bounds for the Gaussian system model (Corollary~\ref{corollary:gauss-outer}) plotted in $(D_s,D_u,r)$ coordinates. The plot contains two surfaces: {\color{blue} the blue surface} shows the bound for Case 1 encoder, and {\color{blue} the red} is the bound for Case 2. For this plot, we set the parameters of the Gaussian semantic source to $P_s = 0.7$, $P_u = 1.0$, $P_{su} = 0.6$; for the wiretap channel: $P = 1$, $P_{N_1} = 0.1$, $P_{N_2} = 0.4$, $\beta_1 = \beta_2 = 1$. We set $R_k = 0$, assuming that no secret key is shared between the encoder and decoder, and secrecy thresholds: $\Delta_s=h(S)$, $\Delta_u=0$, $\Delta_{su}=h(S)$, which correspond to full secrecy for the semantic part of the source.
\ifCLASSOPTIONdraft
\begin{figure} 
\centering\subfigure[Gaussian System Model]{\includegraphics[width=0.35\linewidth,draft=False]{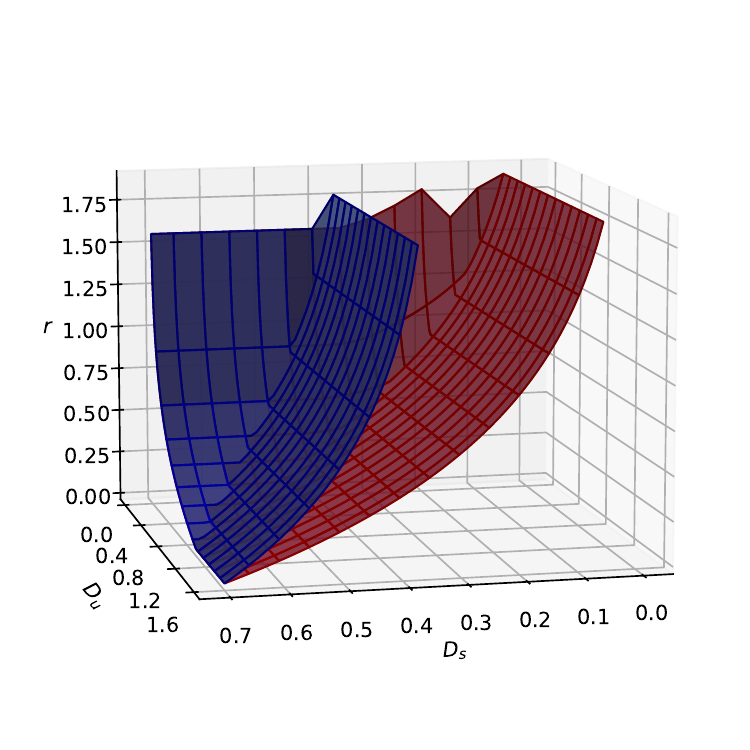}
\label{subfig:gauss-outer-3D}}
\subfigure[Binary System Model]{\includegraphics[width=0.35\linewidth,draft=False]{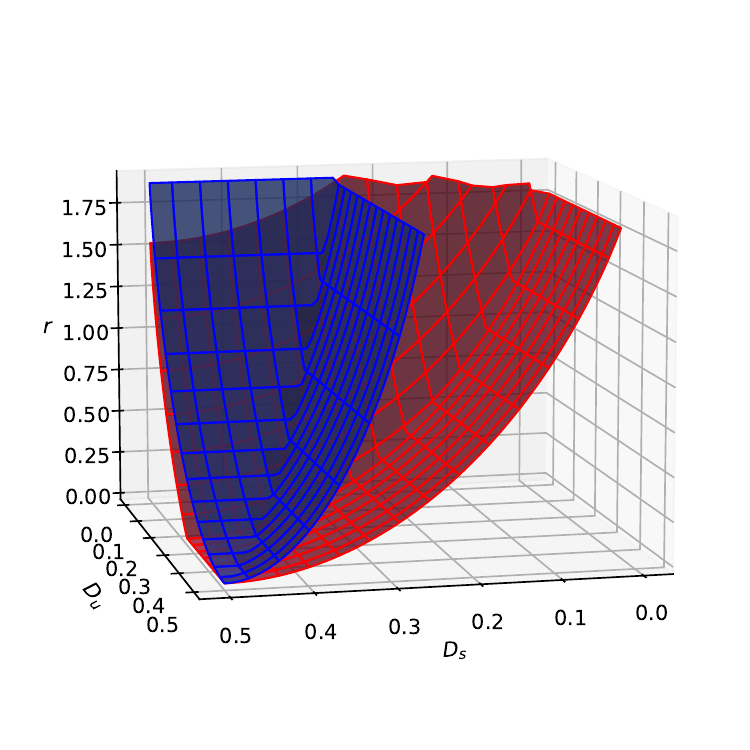}
\label{subfig:binary-outer-3D}}
\caption{Converse bounds the Gaussian and binary system models. {\color{blue} Blue surface:} Case 1 encoder. {\color{blue} Red surface:} Case 2 encoder.}
\label{fig:outer-3D}
\end{figure}
\else
\begin{figure} 
\centering\subfigure[Gaussian System Model]{\includegraphics[width=0.475\linewidth,draft=False]{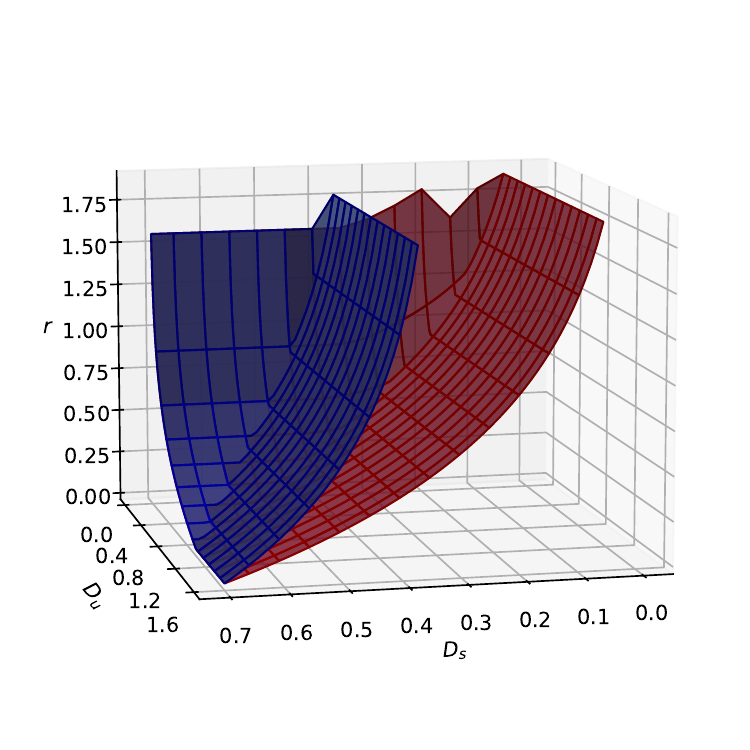}
\label{subfig:gauss-outer-3D}}
\subfigure[Binary System Model]{\includegraphics[width=0.475\linewidth,draft=False]{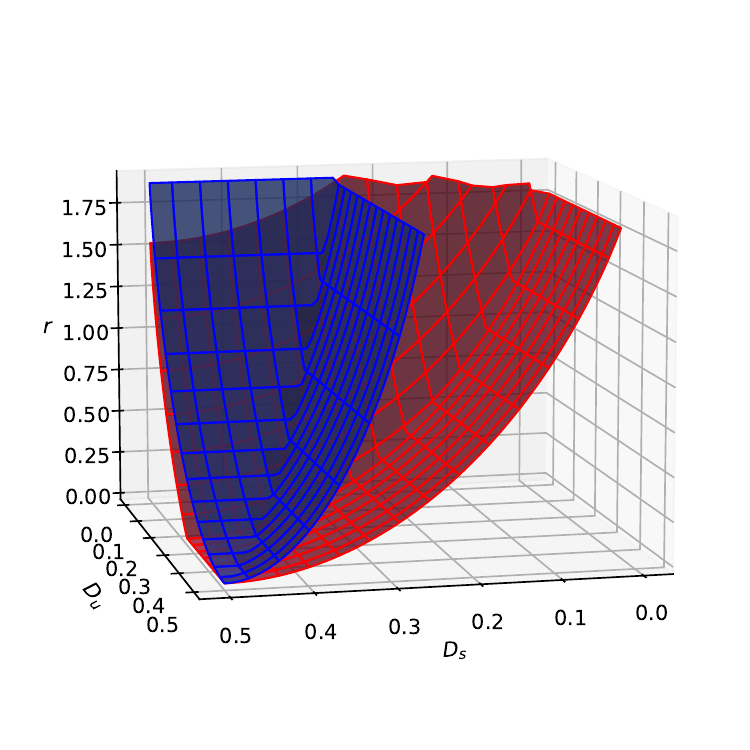}
\label{subfig:binary-outer-3D}}
\caption{Converse bounds the Gaussian and binary system models. {\color{blue} Blue surface:} Case 1 encoder. {\color{blue} Red surface:} Case 2 encoder.}
\label{fig:outer-3D}
\end{figure}
\fi

Meanwhile, for the same parameters, Fig.~\ref{fig:gauss-2D-plots} illustrates the converse and achievability bounds with fixed observation distortion $D_u=0.6$, while varying semantic distortion $D_s$ and $r$. {\color{blue} Fig.~\ref{fig:gauss-2D-plots}, we denote with red} no secrecy case, where $\Delta_s = \Delta_u = \Delta_{su} = 0$. In such case, {\color{blue} red }bounds reflect rate-distortion trade-off, and can serve as a baseline to compare region reduction in non-zero secrecy regime. {\color{blue} Green} curves correspond to the regime of full secrecy on the semantic component only $\Delta_s = \Delta_{su} = h(S)$, $\Delta_u = 0$. Finally, {\color{blue} blue } curves reflect the full secrecy regime where $\Delta_s = h(S)$, $\Delta_u = h(U)$, and $\Delta_u = h(S,U)$.

Note that in Case 1 encoding, there is a minimum threshold distortion $D_s$ for the semantic part (shown as a dashed black line), below which no achievable tuples exist. This is because in Case 1 the encoder can only access the semantic part through the correlation with the observed part.

From Fig~\ref{subfig:outer-case1} and Fig~\ref{subfig:outer-case2} one can see that the focus on protecting only semantic information results in performance gain compared to the scenario when all information is under secrecy constraint. Additionally, Case 2 encoding provides significant gains in performance over Case 1 encoding.

\subsubsection{Inner Bound}
{\color{blue} Although Theorem~\ref{theorem:direct} is formulated with the discrete assumptions for the r.v.s, we can also consider the continuous case via quantization (see \cite[Lemma 3.2]{elgamal2011book} for the general idea).}
Fig.~\ref{subfig:outer-and-inner-case2} presents the comparison of the outer (Corollary~\ref{corollary:gauss-outer}) and inner (Theorem~\ref{theorem:direct}) bounds for the Gaussian system model. In Fig.~\ref{subfig:outer-and-inner-case2}, solid lines denote the outer bounds, while {\color{blue} markers} represent the inner bounds.
These bounds are computed using the Monte Carlo method. Specifically, we consider zero-mean Gaussian random vectors $V_1 = (S,U,\A,\B)$ for the source coding part, and $V_2 = (\ABcc,\ABpp,X,Y,Z)$ for the channel coding side\footnote{Note that $\A=(\Sc,\Sp)$, $\B=(\Uc,\Up)$, $\ABcc = (\Acc,\Bcc)$, and $\ABpp = (\App,\Bpp)$}. We also define functions $\tilde{S} : \A \rightarrow \hat{S}$ and $\tilde{U} : \Sc \times \B \rightarrow \hat{U}$ as minimum mean square error (MMSE) estimators of $S$ and $U$, respectively. In this case $\EX (S - \tilde{S}(\A))^2 = \EX \mathrm{Var} (S|\A)$ and $\EX (U - \tilde{U}(\Sc,\B))^2 = \EX \mathrm{Var} (U|\Sc,\B)$.

We generate randomly proper covariance matrix $\mathbf{\Sigma_1}$ of size $(6 \times 6)$ for vector $V_1$ given fixed sub-covariance matrix $\mathbf{K}$, and the covariance matrix $\mathbf{\Sigma_2}$ of size $(7 \times 7)$ for vector $V_2$, given fixed channel distribution $p_{Y,Z|X}=p_{Y|X}p_{Z|Y}$. These covariance matrices have the following form, where $*$ represent randomly filled values,
\ifCLASSOPTIONdraft
\begin{equation}
\Sigma_1 =
\begin{bmatrix}
P_s & P_{su} & * \\
P_{su} & P_u & * \\
* & * & *
\end{bmatrix},
\qquad
\Sigma_2 =
\begin{bmatrix}
* & * & * & * \\
* & P & P & P \\
* & P & P + P_{N_1} & P + P_{N_1} \\
* & P & P + P_{N_1} & P + P_N
\end{bmatrix}.
\end{equation}
\else
\begin{align}
&\Sigma_1 =
\left[\begin{matrix}
P_s & P_{su} & * \\
P_{su} & P_u & * \\
* & * & *
\end{matrix}\right],
\\
&\Sigma_2 =
\left[\begin{matrix}
* & * & * & * \\
* & P & P & P \\
* & P & P + P_{N_1} & P + P_{N_1} \\
* & P & P + P_{N_1} & P + P_N
\end{matrix}\right].
\end{align}
\fi

Hence, given fixed $\mathbf{\Sigma_1}$ and $\mathbf{\Sigma_2}$ we can numerically compute every term (conditional covariances, entropies, and mutual informations) in our achievability region.
Our simulation involves running calculations for random matrices $\Sigma_1$ and $\Sigma_2$. The simulation forms a scatter of points where we get bound as a minimum for the $r$ axis.


{\color{blue} Fig.~\ref{subfig:outer-and-inner-case2-key} presents comparison of the outer and inner bounds for fixed distortions $D_s = 0.3$ and $D_u = 0.4$, fixed $r = 1.0$, $R_{k2} = 0$, $P = 1$, $P_{n1} = 0.1$, $P_{n2} = 0$ and varying $\Delta_S$ and $R_{k1}$.}

\ifCLASSOPTIONdraft
\begin{figure}
\centering\hfill\subfigure[Outer bound for Case 1 encoding]{\includegraphics[width=0.31\linewidth,draft=False]
{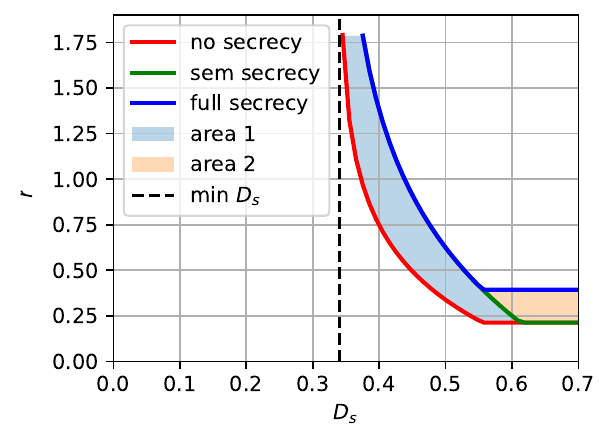}
\label{subfig:outer-case1}}
\centering\hfill\subfigure[Outer bound Case 2 encoding]{\includegraphics[width=0.31\linewidth,draft=False]
{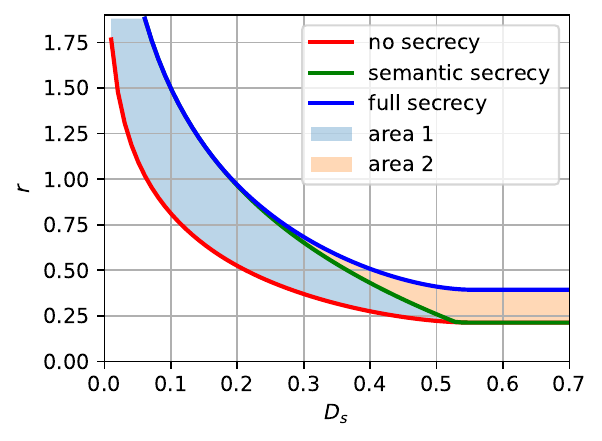}
\label{subfig:outer-case2}}
\centering\hfill\subfigure[Outer and inner bounds]{\includegraphics[width=0.31\linewidth,draft=False]
{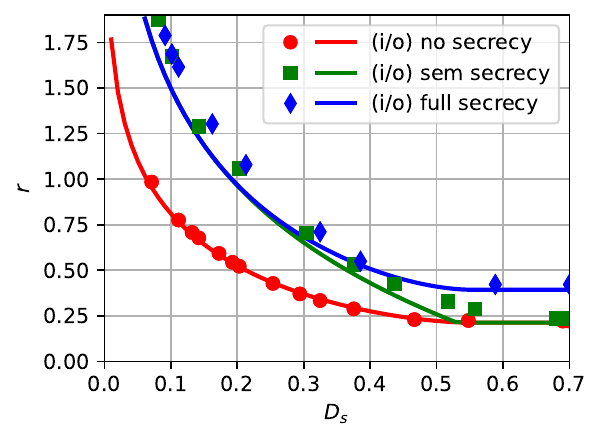}
\label{subfig:outer-and-inner-case2}}
\caption{Inner and outer bounds for the Gaussian system model. {\color{blue} Red} corresponds to the no secrecy case, i.e, $\Delta_s = \Delta_u = \Delta_{su} = 0$. {\color{blue} Green} is to denote full semantic part secrecy case, i.e., $\Delta_s = \Delta_{su} = h(S)$, while $\Delta_u = 0$. {\color{blue} Blue} reflects full secrecy for both parts of the semantic source, i.e., $\Delta_s = h(S)$, $\Delta_u = h(U)$, and $\Delta_{su} = h(S,U)${\color{blue}. The area 1 represents region reduction from no secrecy case if one imposes semantic secrecy constraint, while area 2 shows the gain we get from imposing secrecy only on semantic part of the source comparing to the secrecy constraint on both parts.} In (c), {\color{blue} solid} lines without markers represent outer bounds, while markers -- inner bounds. In the legend on figure (c) the notation (i/o) stands for (inner/outer), respectively.}
\label{fig:gauss-2D-plots}
\end{figure}
\else
\begin{figure}
\centering\hfill\subfigure[Outer bound for Case 1 encoding]{\includegraphics[width=0.475\linewidth,draft=False]
{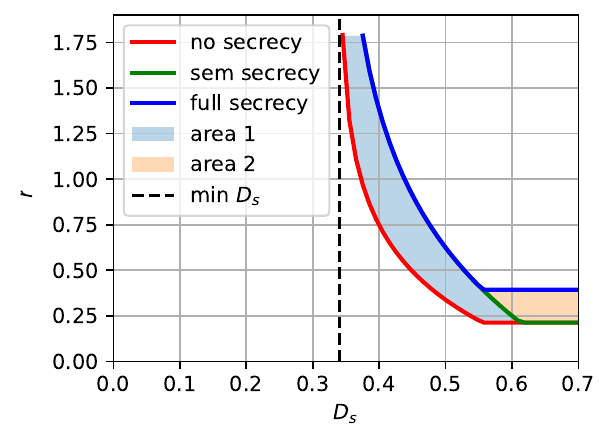}
\label{subfig:outer-case1}}
\centering\hfill\subfigure[Outer bound Case 2 encoding]{\includegraphics[width=0.475\linewidth,draft=False]
{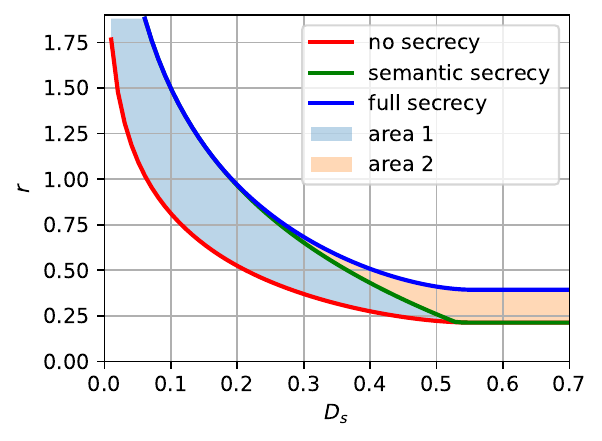}
\label{subfig:outer-case2}}
\centering\hfill\subfigure[Outer and inner bounds]{\includegraphics[width=0.475\linewidth,draft=False]
{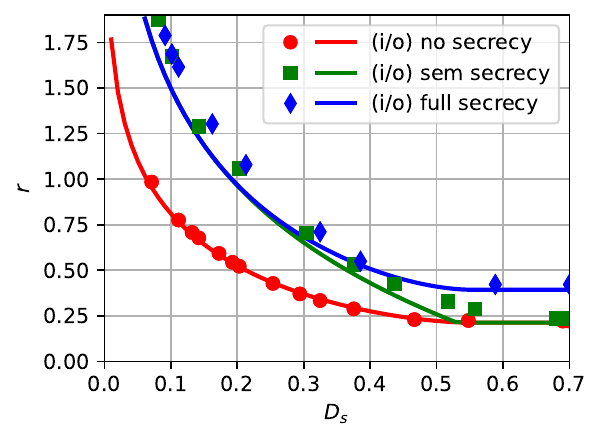}
\label{subfig:outer-and-inner-case2}}
\caption{Inner and outer bounds for the Gaussian system model. {\color{blue} Red} corresponds to the no secrecy case, i.e, $\Delta_s = \Delta_u = \Delta_{su} = 0$. {\color{blue} Green} is to denote full semantic part secrecy case, i.e., $\Delta_s = \Delta_{su} = h(S)$, while $\Delta_u = 0$. {\color{blue} Blue} reflects full secrecy for both parts of the semantic source, i.e., $\Delta_s = h(S)$, $\Delta_u = h(U)$, and $\Delta_{su} = h(S,U)${\color{blue}. The area 1 represents region reduction from no secrecy case if one imposes semantic secrecy constraint, while area 2 shows the gain we get from imposing secrecy only on semantic part of the source comparing to the secrecy constraint on both parts.} In (c), {\color{blue} solid} lines without markers represent outer bounds, while markers -- inner bounds. In the legend on figure (c) the notation (i/o) stands for (inner/outer), respectively.}
\label{fig:gauss-2D-plots}
\end{figure}
\fi
\subsection{Binary System Model}
\label{sub:binary-model}

We next consider the fundamental limits of the transmission of binary semantic sources over binary symmetric wiretap channels.

Consider a binary semantic source $(S,U)$ where: semantic part is distributed as $S \sim \mathcal{B}(0.5)$, and the observed part $U$ is received through the binary symmetric channel (BSC) with crossover probability $\alpha$, where $\alpha \in [0,0.5]$.
\begin{equation}
    S \rightarrow \text{BSC}(\alpha) \rightarrow U
\end{equation}
We set the Hamming distortion measures for both parts of the source:
$d_s(x,\hat{x}) = d_u(x,\hat{x}) = \mathbb{I}_{\{x \neq \hat{x}\}}$. The channel is modelled as the BSC given {\color{blue}$0 \leq \epsilon_1 \leq \epsilon_2 \leq 0.5$}:
\begin{align*}
\text{Main Channel: } & Y = X \oplus N_1, \quad N_1 \sim \mathcal{B}(\epsilon_1) \\
\text{Wiretap Channel: } & Z = {\color{blue} X} \oplus N_2, \quad N_2 \sim \mathcal{B}(\epsilon_2)
\end{align*}


\begin{corollary}[Binary Converse Bound] \label{prop:binary-outer}
For the binary system model defined in Subsection~\ref{sub:binary-model}, any achievable tuple satisfies:
\ifCLASSOPTIONdraft
\begin{align}
R(D_s,D_u) &\leq r(1 - H_b(\epsilon_1)) \\
\Delta_s &\leq R_k + r [H_b(\gamma_1 * \epsilon_2) - H_b( \gamma_1 * \epsilon_1)] + 1 - R_s(D_s) \\
\Delta_u &\leq R_k + r [H_b(\gamma_2 * \epsilon_2) - H_b( \gamma_2 * \epsilon_1)] + H_b(\alpha) - R_u(D_u) \\
\Delta_{su} &\leq R_k + r [H_b(\epsilon_2) - H_b(\epsilon_1)] + H_b(\alpha) + 1 - R(D_s,D_u)
\end{align}
\else
\begin{align}
&R(D_s,D_u) \leq r(1 - H_b(\epsilon_1)) \\
&\Delta_s \leq r [H_b(\gamma_1 {*} \epsilon_2) - H_b( \gamma_1 {*} \epsilon_1)] \nonumber \\ &\quad + 1 - R_s(D_s) + R_k \\
&\Delta_u \leq r [H_b(\gamma_2 {*} \epsilon_2) - H_b( \gamma_2 {*} \epsilon_1)] \nonumber \\ &\quad 
+ H_b(\alpha) - R_u(D_u) + R_k \\
&\Delta_{su} \leq r [H_b(\epsilon_2) - H_b(\epsilon_1)]  \nonumber \\
&\quad + H_b(\alpha) + 1 - R(D_s,D_u) + R_k
\end{align}
\fi
for some $\gamma_1, \gamma_2 \in [0,1]$, in Case 1 encoding $\gamma_2 = 0$.
\end{corollary}
\begin{remark}
    For simplicity, one can always set $\gamma_1 = \gamma_2 = 0$ and obtain a valid converse bound.
\end{remark}

\subsubsection{Proof of Corollary~\ref{prop:binary-outer}}
This corollary follows from the general converse (Theorem~\ref{theorem:converse}). Note that the proof for general converse continues to hold for the binary system model. The only difference is that we can derive explicit equations for some terms in the converse bound. For instance, we have the following bound $I(X;Y) \leq 1 - H_b(\epsilon_1)$, which {\color{blue} corresponds} to channel capacity of the BSC. Similarly, the secrecy capacity of the binary wiretap channel is {\color{blue} $I(\ABpp;Y) - I(\ABpp;Z) \leq H_b(\epsilon_2) - H_b(\epsilon_1)$. We also have $I(\App;Y) - I(\App;Z) \leq H_b(\gamma_1 * \epsilon_2) - H_b( \gamma_1 * \epsilon_1)$, and $I(\Bpp;Y) - I(\Bpp;Z) \leq H_b(\gamma_2 * \epsilon_2) - H_b(\gamma_2 * \epsilon_1)$} for some $\gamma_1, \gamma_2 \in [0,1]$, where in Case 1 encoding $\gamma_2 = 0$, with the assumption that $X = \App \oplus \mathcal{B}(\gamma_1)$ and $X = \Bpp \oplus \mathcal{B}(\gamma_2)$.

\subsubsection{Rate-Distortion Functions}
Closed-form solutions for RDFs in binary the converse depend on the encoder case and are presented in Lemma~\ref{lemma:marginal-binary-rdfs} and Lemma~\ref{lemma:joint-binary-rdfs}.

\begin{lemma}[Marginal Binary RDFs] \label{lemma:marginal-binary-rdfs}
Marginal RDFs for the Bernoulli distributed semantic source have the following closed form solutions depending on the encoder case. \\
\textbullet \  Case 1 and Case 2 RDF for the observation \cite{elgamal2011book}:
\begin{equation}
R_u(D_u) = \begin{cases} 
H_b(\alpha) - H_b(D_u), & D_u \leq \alpha \\
0, & \text{otherwise}
\end{cases}
\end{equation} \\
\textbullet \ Case 2 RDF for the semantic part \cite{elgamal2011book}:
\begin{equation}
R_{s,d}(D_s) = \begin{cases} 
1 - H_b(D_s), & D_s \leq 0.5 \\
0, & \text{otherwise}
\end{cases}
\end{equation} \\
\textbullet \ Case 1 RDF for the semantic part \cite[Problem 3.8]{berger2003}:
\begin{equation}
R_{s,i}(D_s) = \begin{cases} 
1 - H_b\left(\frac{D_s - \alpha}{1 - 2\alpha}\right), & \alpha \leq D_s < 0.5 \\
+\infty, & D_s < \alpha \\
0, & D_s \geq 0.5
\end{cases}
\end{equation}
\end{lemma}

\begin{lemma}[Joint Binary RDFs]
\label{lemma:joint-binary-rdfs}
Joint RDFs for the Bernoulli distributed semantic source for Case 1 and Case 2 encoder can be written as follows. \\
\textbullet \ Case 1 Binary RDF \cite{stavrou2023}:
\begin{equation}
R_i(D_s,D_u) = \max\{ R_u(D_u), R_{s,i}(D_s)\},
\end{equation}
where $R_u(D_u)$ and $R_{s,i}(D_s)$ are given in Lemma~\ref{lemma:marginal-binary-rdfs}. \\
\textbullet \ Case 2 Binary RDF can be found in \cite[Theorem 2]{nayak2007}.
\end{lemma}

\subsection{Numerical Analysis of the Binary System Model} 
\label{sub:binary-numerics}
Fig.~\ref{subfig:binary-outer-3D} shows 3D plot of the converse bound of Corollary~\ref{prop:binary-outer}. The parameters used for this plot are: $R_k = 0$, $\Delta_s = H_b(0.5)$, $\Delta_u = 0$, $\Delta_{su} = H_b(0.5)$, corresponding to full secrecy for the semantic source component, $\alpha = 0.25$, $\epsilon_1 = 0.1$, {\color{blue} $\epsilon_2 = 0.34$}, and $\gamma_1 = \gamma_2 = 0$. The tradeoff between fidelity $D_s$ and secrecy $\Delta_s$ for a fixed $r = 1$ and $D_u=0.25$ presented in Fig.~\ref{subfig:outer-2D-binary-case1-case2-Rk}. In Fig.~\ref{subfig:outer-2D-binary-case1-case2-Rk}, we consider {\color{blue} different $\epsilon_2 = 0.11$} and two secrecy-key rates $R_k=0$ and $R_k=0.1$. From Fig.~\ref{subfig:outer-2D-binary-case1-case2-Rk} one can see that equivocation $\Delta_s$ has saturation point at some distortion $D_s$, so that further compression of the semantic part does not contribute to the semantic part secrecy.
\ifCLASSOPTIONdraft
\begin{figure}
\centering\subfigure[{\color{blue} Outer bounds for the binary system model.}]{\includegraphics[width=0.31\linewidth,draft=False]
{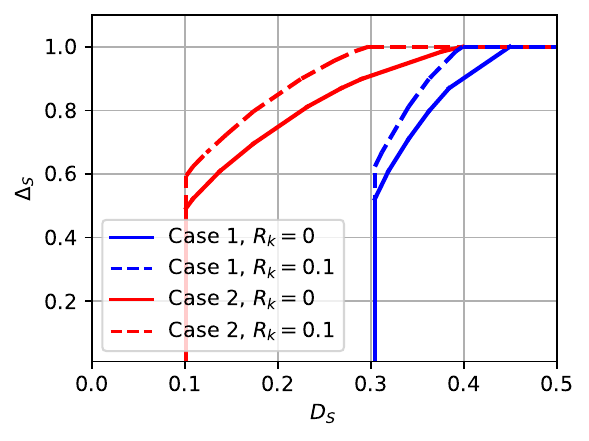}
\label{subfig:outer-2D-binary-case1-case2-Rk}}
\centering\subfigure[{\color{blue} Outer and Inner bounds for the Gaussian system model.}]{\includegraphics[width=0.31\linewidth,draft=False]
{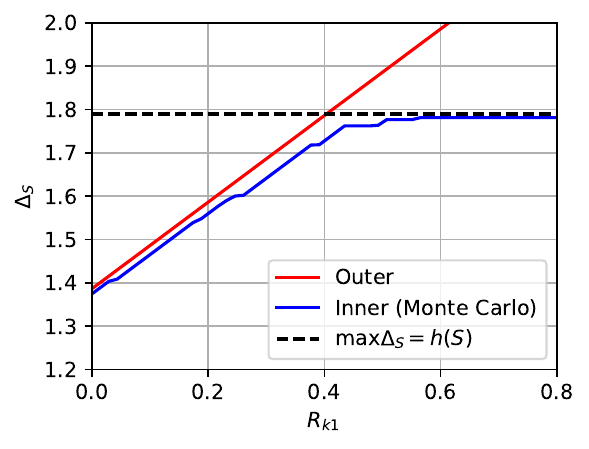}
\label{subfig:outer-and-inner-case2-key}}
\caption{\color{blue}. The effect of the secret key rate on the achievable region.}
\label{fig:key}
\end{figure}
\else
\begin{figure}
\centering\subfigure[{\color{blue} Outer bounds for the binary system model.}]{\includegraphics[width=0.475\linewidth,draft=False]
{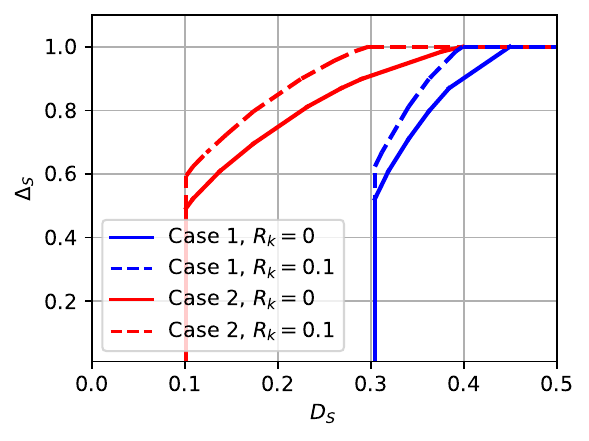}
\label{subfig:outer-2D-binary-case1-case2-Rk}}
\centering\subfigure[{\color{blue} Outer and Inner bounds for the Gaussian system model.}]{\includegraphics[width=0.475\linewidth,draft=False]
{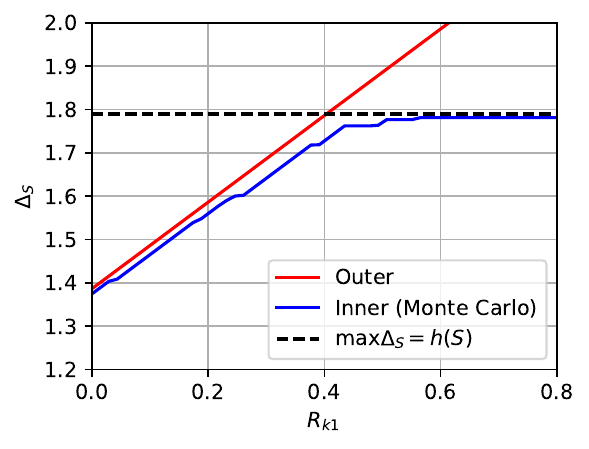}
\label{subfig:outer-and-inner-case2-key}}
\caption{\color{blue} The effect of the secret key rate on the achievable region.}
\label{fig:key}
\end{figure}
\fi
\color{teal} \section{Concluding Remarks}
\label{sec:conclusion}
This work presents an information-theoretic characterisation of the secure semantic-aware communication. The characterisation includes the converse bound and the achievability scheme. The converse region explicitly contains rate-distortion functions, which are easy to evaluate numerically, as they have closed-form solutions for some common distributions. The achievability scheme relies on the novel four-layer superposition code, where the private parts of the message are superposed on the public ones with separation for semantic and observed parts of the source. Such approach allows us to analyse the equivocation separately for each part.

We investigated two cases of source output, one of which differs in the extended information provided to the encoder, that is, samples of the semantic information. The numerical results for Gaussian and binary system models show that if the encoder has direct access to the semantic part, it can perform better. Additionally, the numerical results illustrate that our achievability bound is tight with the converse bound in no secrecy case for the Gaussian system model. {\color{teal} In full semantic secrecy and full joint secrecy scenarios there is gap between converse and achievability bounds. Future research may include proposal of the achievability scheme that is less complex than the current one and is tight with the converse bound given Remark~\ref{remark:converse}. Additionally,} the open question whether separation of source {\color{blue} coding, encryption,} and channel coding {\color{teal} is asymptotically optimal} for this model.
\section*{Acknowledgements}
This research is supported by the 5/6GIC, Institute for Communication Systems (ICS), University of Surrey. The first author of this work also thanks the author of \cite{li2021} for providing the PSITIP software, which facilitated routine checks of information-theoretic quantities.

\bibliographystyle{IEEEtran}  
\bibliography{main}

\newpage
\clearpage
\begin{center}
{\LARGE Supplementary Material}
\end{center}
\begin{appendices}
    {\color{blue}
\section{Proof of Cardinality Bounds for $\App$ and $\Bpp$}
\label{appendix:cardinality}

The proof of the cardinality of $\App,\Bpp$ involves standard cardinality bounding techniques. Here we outline the proof for the cardinality of $\App$, the proof for $\Bpp$ follows similar steps. Firstly, consider $\App'$ of arbitrary cardinality such that $\App' \to X \to (Y,Z)$. We expand $I(\App';Y) - I(\App';Z)$ as follows:
\ifCLASSOPTIONdraft
\begin{align}
    &I(\App';Y) - I(\App';Z) = H(Y) - H(Y|\App') - H(Z) + H(Z|\App') \\
    &= \sum_{w'_1 \in \App'} p_{\App'}(w'_1) \left( H(Y) - H(Y|\App' = w'_1) - H(Z) + H(Z|\App' = w'_1) \right)
\end{align}
\else
\begin{align}
    &I(\App';Y) - I(\App';Z) \\
    &= H(Y) - H(Y|\App') - H(Z) + H(Z|\App') \\
    &= \sum_{w'_1 \in \App'} p_{\App'}(w'_1) ( H(Y) - H(Y|\App' = w'_1) \nonumber \\
    &\quad - H(Z) + H(Z|\App' = w'_1) )
\end{align}
\fi
Terms $H(Y)$ and $H(Z)$ are functions of $p_X$. Consider the following vectors:
\ifCLASSOPTIONdraft
\begin{align}
    &\mathbf{v} = \left( I(\App';Y) - I(\App';Z), p_X(1), ..., p_X(|\mathcal{X}| - 1) \right), \\
    &\mathbf{v}_{w'_1} {=} \left( H(Y) {-} H(Y|w'_1) {-} H(Z) {+} H(Z|w'_1), p_{X|\App'}(1|w'_1),..., p_{X|\App'}(|\mathcal{X}| - 1 | w'_1) \right)
\end{align}
\else
\begin{align}
    &\mathbf{v} = \left( I(\App';Y) - I(\App';Z), p_X(1), ..., p_X(|\mathcal{X}| - 1) \right), \nonumber \\
    &\mathbf{v}_{w'_1} {=} ( H(Y) {-} H(Y|w'_1) {-} H(Z) {+} H(Z|w'_1), \nonumber \\
    &\quad p_{X|\App'}(1|w'_1),..., p_{X|\App'}(|\mathcal{X}| - 1 | w'_1) )
\end{align}
\fi
Note that we only need vector till $p_X(|\mathcal{X}| - 1)$ to fully characterise distribution $p_X$ as the last element can be derived given that $\sum_{x \in \mathcal{X}} p_X(x) {=} 1$. Also $p_{X}(x) = \sum_{w'_1 \in \mathcal{W}'_1} p_{\App'}(w'_1) p_{X|\App'}(x|w'_1)$. The vector $\mathbf{v}$ is of size $|\mathcal{X}|$ and is a convex combination of $\mathbf{v}_{w'_1}$: $\mathbf{v} = \sum_{w'_1 \in \App'} p_{\App'}(w'_1) \mathbf{v}_{w'_1}$. Then we apply Fenchel-Eggleston-Carathéodory's Theorem \cite{elgamal2011book} and we use equivalent of $\App'$ and call it $\App$ with cardinality $|\mathcal{W}_1| \leq |\mathcal{X}|$ to equivalently characterise vector $\mathbf{v}$.
}
    {\color{blue}
\section{Codebook Generation}
\label{appendix:codebook}
}

This appendix explains in detail codebook generation for the achievability scheme.

\subsection{Source Coding Codebook}

Let $\RSc$, $\RSp$, $\RUc$, and $\RUp$ be a non-negative rates.
Also, let $\A = (\Sc,\Sp)$ and $\B = (\Uc,\Up)$ be auxiliary r.v.s. {\color{blue}for the codebook generation with arbitrary joint distribution in Case 2 encoding, and given that $S \to U \to (\A,\B)$ in Case 1 encoding}.
We define the following codebook for the source encoding stage; see Fig.~\ref{subfig:src-codebook}.
\begin{enumerate}
    \item Independently uniformly at random pick\footnote{Further, for brevity, we use word ``pick" alone and we mean independently uniformly at random.} $2^{k\RSc}$ sequences of length $k$ from $\T^k(\Sc)$ and call it $\sck$. Then, for each $\sck$ pick $2^{k\RSp}$ sequences of length $k$ from $\T^k(\Sp|\sck)$ and call it $\spk$, where $i \in [1,...,2^{k\RSc}]$ and $j \in [1,...,2^{k\RSp}]$.
    \item For each $\sck$ pick $2^{k\RUc}$ sequences of length $k$ from $\T^k(\Uc|\sck)$ and call it $\uck$. Then, for each $\uck$ pick $2^{k\RUp}$ sequences of length $k$ from $\T^k(\Up|\sck,\uck)$ and call it $\upk$, where $l \in [1,...,2^{k\RUc}]$ and $p \in [1,...,2^{k\RUp}]$.
\end{enumerate}

In order to pick the required number of typical sequences for this codebook, the following rate conditions must be satisfied, which follows from the property for the size of the typical set.
\begin{align}
\label{ineq:rate-Ri-Rj-entropy}
    \RSc &< H(\Sc) - \delta_{\color{blue} k}, \ \quad \RSp < H(\Sp|\Sc) - \delta_{\color{blue} k}, \\
\label{ineq:rate-Rl-Rp-entropy}
    \RUc &< H(\Uc|\Sc) - \delta_{\color{blue} k}, \ \RUp < H(\Up|\Sc,\Uc) - \delta_{\color{blue} k},
\end{align}
where $\delta_{\color{blue} k} \to 0$ with $k \to \infty$.

\subsection{Channel Codebook}
Let $\RAcc$, $\RApp$, $\RBcc$, and $\RBpp$ be non-negative rates. Additionally, let $\ABcc = (\Acc,\Bcc)$ and $\ABpp = (\App,\Bpp)$ be auxiliary r.v.s. {\color{blue} such that $(\ABcc,\ABpp) \to X \to (Y,Z)$}. Also, let $\Rms, \Rmu$ be non-negative rates.
We generate channel codebook as follows, which is illustrated in Fig.~\ref{subfig:ch-codebook}:
\begin{enumerate}
    \item Pick $2^{k\RAcc}$ sequences of length $n$ from $\T^n(\Acc)$ and call it $\qsn$, where $i' \in [1,...,2^{k\RAcc}]$.
    \item For each $\qsn$ pick $2^{k(\RApp + \Rms)}$ sequences of length $n$ from $\T^n(\App|\qsn)$ and call it $\wsn$, where $j' \in [1,...,2^{k\RApp}]$ and $\ws \in [1,...,2^{k\Rms}]$.
    \item For each $\qsn$ pick $2^{k\RBcc}$ sequences of length $n$ from $\T^n(\Bcc|\qsn)$ and call it $\qun$, where $l' \in [1,...,2^{k\RBcc}]$.
    \item For each $\qun$ pick $2^{\color{blue} k(\RBpp + \Rmu)}$ sequences of length $n$ from $\T^n(\Bpp|\qsn,\qun)$ and call it $\wun$, where $p' \in [1,...,2^{k\RBpp}]$ and $\wu \in [1,...,2^{k\Rmu}]$.
\end{enumerate}



{\color{blue} Final step includes for each $\qsns,\quns,\wsns,\wuns$ generation of $x^n$ such that $p(x^n|\qsns,\quns,\wsns,\wuns) = \prod_{i=1}^{n} p(x_i|q_{1,i},q_{2,i},w_{1,i},w_{2,i})$}.

In order to have a sufficient number of typical sequences, the rates should satisfy the following conditions: 
\begin{align}
\label{ineq:ch-cb-Ws-rate}
    \RAcc &< (r + \epsilon) H(\Acc) - \delta_{\color{blue} k}, \\
\label{ineq:ch-cb-Wu-rate}
    \RBcc &< (r + \epsilon) H(\Bcc|\Acc) - \delta_{\color{blue} k}, \\
\label{ineq:ch-cb-Qu-rate}
    \RApp + \Rms &< (r + \epsilon) H(\App|\Acc) - \delta_{\color{blue} k}, \\
\label{ineq:ch-cb-Qs-rate}
    \RBpp + \Rmu &< (r + \epsilon) H(\Bpp|\Acc,\Bcc) - \delta_{\color{blue} k},
\end{align}
{\color{blue} where $\delta_{\color{blue} k} \to 0$ with $k \to \infty$.}
    {\color{blue}
\section{Definition of the Error Event}
\label{appendix:errors}

Our achievability assumes that errors can occur at several stages of our coding scheme. We define the total error event $\mathcal{E}$ as the event where the sequences are not jointly typical, the encoder fails to encode, and the receiver fails to correctly recover the transmitted indices.
\ifCLASSOPTIONdraft
\begin{equation}
    \mathcal{E} = \{ \text{Joint Typicality} \} \cup \{ \text{Encoding Failure} \} \cup \{ \text{Decoding Failure} \} \doteq \bigcup_{i=1}^{10} \Err_i,
\end{equation}
\else
\begin{align}
    \mathcal{E} &= \{ \text{Joint Typ.} \} \cup \{ \text{Enc. Failure} \} \cup \{ \text{Dec. Failure} \} \nonumber \\
    &\doteq \bigcup_{i=1}^{10} \Err_i,
\end{align}
\fi
where each $\Err_i$ is defined below. We upper-bound probability of an error event $\Err$ using union bound:
\begin{equation*}
    P_{\Err} \doteq Pr \{\Err\} \leq  \sum_{i=1}^{10} \Pr\{\Err_i\},
\end{equation*}

\subsection{Joint Typicality Events}
\begin{enumerate}
    \item $\Err_1 \doteq \{ (S^k, U^k, \A^k, \B^k) \not\in \T^k(S,U,\A,\B) \}$
    \item $\Err_2 \doteq \{ (\ABcc^n,\ABpp^n, Y^n) \not\in \T^n(\ABcc,\ABpp, Y) \}$
\end{enumerate}
Given $k \to \infty$ and $n \to \infty$ the probabilities of $\Err_1$ and $\Err_2$ go to zero due to asymptotic equipartition property (AEP).

\subsection{Source Encoding Errors}
Source encoding errors are events when there is no jointly typical sequence to pick from a typical set.
\ifCLASSOPTIONdraft
\begin{enumerate}
    \item $\Err_3 \doteq \{ \not\exists i : (\sck, v^k) \in \T^k(\Sc, V) \}$
    \item $\Err_4 \doteq \{ \not\exists j : (\spk, v^k) \in \T^k(\Sp, V | \sck) \}$
    \item $\Err_5 \doteq \{ \not\exists l : (\uck, v^k) \in \T^k(\Uc, V|\sck) \}$
    \item $\Err_6 \doteq \{ {\not\exists} p : (\upk, v^k) \in \T^k(\Up, V | \sck,\uck) \}$
\end{enumerate}
\else
\begin{align*}
    &1) \ \Err_3 \doteq \{ \not\exists i : (\sck, v^k) \in \T^k(\Sc, V) \} \\
    &2) \ \Err_4 \doteq \{ \not\exists j : (\spk, v^k) \in \T^k(\Sp, V | \sck) \} \\
    &3) \ \Err_5 \doteq \{ \not\exists l : (\uck, v^k) \in \T^k(\Uc, V|\sck) \} \\
    &4) \ \Err_6 \doteq \{ {\not\exists} p : (\upk{,} v^k) {\in} \T^k(\Up{,} V | \sck{,}\uck) \}
\end{align*}
\fi
Given rate conditions (\ref{ineq:rate-Ri})-(\ref{ineq:rate-Rp}) and $k \to \infty$, probabilities of events $\Err_3,...,\Err_6$ go to zero.

\subsection{Decoding Errors}
At the decoding stage at the legitimate receiver an error occurs if there is no unique codeword jointly typical with the channel sequence $y^n$.
\ifCLASSOPTIONdraft
\begin{enumerate}
    \item $\Err_7 \doteq 
    \{
    \exists i'_1 \neq i'_2 :
    ( q^n_1(i'_1), q^n_1(i'_2) ) \in \T^n(\Acc,Y)
    \},$
    \item $\Err_8 \doteq 
    \{
    \exists i'_1 \neq i'_2, l'_1 \neq l'_2 :
    ( q^n_2(i'_1,l'_1), q^n_2(i'_2,l'_2) ) \in \T^n(\Bcc,Y|\Acc)
    \},$
    \item $\Err_9 \doteq 
    \{
    \exists i'_1 \neq i'_2, j'_1 \neq j'_2, m'_1 \neq m'_2 :
    ( w^n_1(i'_1,l'_1,m'_1), w^n_1(i'_2,l'_2,m'_2) ) \in \T^n(\App,Y|\Acc)
    \},$
    \item $\Err_{10} \doteq 
    \{
    \exists i'_1 \neq i'_2, l'_1 \neq l'_2, p'_1 \neq p'_2, m'_1 \neq m'_2 :
    ( w^n_2(i'_1,l'_1,p'_1,m'_1), w^n_2(i'_2,l'_2,p'_2,m'_2) ) \in \T^n(\Bpp,Y|\Acc,\Bcc)
    \}.$
\end{enumerate}
\else
\begin{align*}
    &1) \ \Err_7 \doteq 
    \{
    \exists i'_1 \neq i'_2 :
    ( q^n_1(i'_1), q^n_1(i'_2) ) \in \T^n(\Acc,Y)
    \}, \\
    &2) \ \Err_8 \doteq 
    \{
    \exists i'_1 \neq i'_2, l'_1 \neq l'_2 : \\
    &\qquad 
    ( q^n_2(i'_1,l'_1), q^n_2(i'_2,l'_2) ) \in \T^n(\Bcc,Y|\Acc)
    \}, \\
    &3) \ \Err_9 \doteq 
    \{
    \exists i'_1 \neq i'_2, j'_1 \neq j'_2, m'_1 \neq m'_2 : \\
    &\qquad 
    ( w^n_1(i'_1,l'_1,m'_1), w^n_1(i'_2,l'_2,m'_2) ) \in \T^n(\App,Y|\Acc)
    \}, \\
    &4) \ \Err_{10} \doteq 
    \{
    \exists i'_1 \neq i'_2, l'_1 \neq l'_2, p'_1 \neq p'_2, m'_1 \neq m'_2 : \\
    &\qquad 
    ( w^n_2(i'_1,l'_1,p'_1,m'_1), w^n_2(i'_2,l'_2,p'_2,m'_2) )  \\
    &\qquad \in \T^n(\Bpp,Y|\Acc,\Bcc)
    \}.
\end{align*}
\fi
Given rate conditions (\ref{ineq:rate-R'i})-(\ref{ineq:rate-R'p+R'm2}) and $k \to \infty$, probabilities of events $\Err_7,...,\Err_{10}$ go to zero.
}
    {\color{blue}
\section{Equivocation Inequality}
\label{appendix:eqv-ineq}
In this Appendix we prove the key inequality used in the equivocation analysis for achievability scheme.
}
Specifically, we show that for arbitrary $S$, $Z$, $W$, and $C=(C_0,C_1)$ the following inequality holds.
\ifCLASSOPTIONdraft
\begin{align}
    H(S) {\color{blue} {+} H(C_0|S)} {+} H(W|C_0) {+} H(Z|W, C) \leq H(S|Z) {+} H(C) {+} H(W|C) {+} H(Z|C_0) {+} I(Z; S|C)
    \label{ineq:eqv-inter}
\end{align}
\else
\begin{align}
    &H(S) {\color{blue} {+} H(C_0|S)} {+} H(W|C_0) {+} H(Z|W, C) \\
    &\leq H(S|Z) {+} H(C) {+} H(W|C) {+} H(Z|C_0) {+} I(Z; S|C)
    \label{ineq:eqv-inter}
\end{align}
\fi

Then with a straightforward rearrangement we can obtain the inequality in the initial form with $H(S|Z)$ on the left-hand side. The proof of (\ref{ineq:eqv-inter}) is as follows.
\ifCLASSOPTIONdraft
\begin{align}
&H(S) {\color{blue} + H(C_0|S)} + H(W|C_0) + H(Z|C_0, C_1, W)\\
&\leq H(S,C_0) + H(W, Z, C_1|C_0)\\
&\leq H(S|C_0)+H(Z|C_0, C_1)+H(C_0, C_1, W)\\
&\leq I(S; Z|C_0, C_1)+H(S, Z|C_0)+H(C_0, C_1, W)\\
&\leq H(S|Z)+H(Z|C_0)+I(S; Z|C_0, C_1)+H(C_0, C_1, W).
\end{align}
\else
\begin{align*}
&H(S) {\color{blue} + H(C_0|S)} + H(W|C_0) + H(Z|C_0, C_1, W) \nonumber \\
& \leq H(S,C_0) + H(W, Z, C_1|C_0)\\
& \leq H(S|C_0) + H(Z|C_0, C_1) + H(C_0, C_1, W)\\
& \leq I(S; Z|C_0, C_1) + H(S, Z|C_0) + H(C_0, C_1, W)\\
& \leq H(S|Z) + H(Z|C_0) + I(S; Z|C_0, C_1) \\
&\quad + H(C_0, C_1, W).
\end{align*}
\fi 
    {\color{blue}
\section{Equivocation Mutual Information Terms Analysis}
\label{appendix:mi-term}

This appendix provides single-letter upper bounds for $I(S^k;Z^n|I',J')$ and $I(Z^n;U^k|I',L',P')$, which appear in the equivocation analysis in Subsection~\ref{subsubsec:eqv-analysis}.
\ifCLASSOPTIONdraft
\begin{align}
    &I(S^k;Z^n|I',J') \stackrel{(a)}{=} I(S^k;Z^n|\hat{I},J) \stackrel{(b)}{=} I(S^k;Z^n|I,J,K_1) {+} I(S^k;K_1|\hat{I},J) {-} I(S^k;K_1|Z^n,\hat{I},J) \nonumber \\
    & = I(S^k;Z^n|I,J) + I(S^k;K_1|Z^n,I,J) - I(S^k;K_1|I,J) + I(S^k;K_1|\hat{I},J) - I(S^k;K_1|Z^n,\hat{I},J) \nonumber \\
    &\leq^{(c)} I(S^k;Z^n|I,J) + H(K_1|Z^n,I,J) + H(K_1) - I(I;K_1|\hat{I}) \nonumber \\
    &\leq^{(d)} I(S^k;Z^n,T|I,J) = I(S^k;T|I,J) + I(S^k;Z^n|I,J,T),
    \label{eq:nasty-mi-term}
\end{align}
\else
\begin{align}
    &I(S^k;Z^n|I',J') \stackrel{(a)}{=} I(S^k;Z^n|\hat{I},J) \\
    &\stackrel{(b)}{=} I(S^k;Z^n|I,J,K_1) + I(S^k;K_1|\hat{I},J) \nonumber \\ &\quad
    - I(S^k;K_1|Z^n,\hat{I},J) \\
    & = I(S^k;Z^n|I,J) + I(S^k;K_1|Z^n,I,J) \nonumber \\ &\quad
    - I(S^k;K_1|I,J) + I(S^k;K_1|\hat{I},J) \nonumber \\ &\quad
    - I(S^k;K_1|Z^n,\hat{I},J) \\
    &\leq^{(c)} I(S^k;Z^n|I,J) + H(K_1|Z^n,I,J) + H(K_1) \nonumber \\ &\quad
    - I(I;K_1|\hat{I}) \\
    &\leq^{(d)} I(S^k;Z^n,T|I,J) = I(S^k;T|I,J) \nonumber \\ &\quad
    + I(S^k;Z^n|I,J,T),
    \label{eq:nasty-mi-term}
\end{align}
\fi
where (a) is due to $g^{-1}_2$ mapping, (b) is because we can decrypt $I$ given $\hat{I}$ and $K_1$, (c) is because $I(S^k;K_1|Z^n,\hat{I},J) = I(I;K_1|\hat{I})$, (d) is due to $H(K_1|Z^n,I,J) \leq \epsilon$\footnote{\color{blue} We assume that the eavesdropper can decode $\hat{I}$ with the small probability of error $H(K_1|Z^n,I,J) \leq  H(K_1|\hat{I},I) + \epsilon = \epsilon$.} and $I(I;K_1|\hat{I}) = H(K_1)$, for arbitrary $T$. We set $T = (I,J,L,P)$ then the first term of (\ref{eq:nasty-mi-term}):
\ifCLASSOPTIONdraft
\begin{align}
    I(S^k;I,J,L,P|I,J) =^{(a)} I(S^k;\B^k|\A^k) \leq^{(b)} k I(S;\B|\A) + \epsilon,
    \label{eq:nasty-mi-derivation}
\end{align}
\else
\begin{align}
    &I(S^k;I,J,L,P|I,J) =^{(a)} I(S^k;\B^k|\A^k) \\
    &\leq^{(b)} k I(S;\B|\A) + \epsilon,
    \label{eq:nasty-mi-derivation}
\end{align}
\fi
where (a) is because $\A^k$ and $\B^k$ are functions of the corresponding indices, (b) is valid for $k \to \infty$ due to the properties of the typical sequences given that $(S^k,\A^k,\B^k) \in \T^k(S,\A,\B)$, which in Case 2 holds directly and in Case 1 encoding is due to Markov Lemma \cite{elgamal2011book} given that $S \to U \to (\A,\B)$.
The second term of (\ref{eq:nasty-mi-term}) is $I(S^k;Z^n|I,J,L,P) = 0$ due to $(S^k,U^k) \to (I,J,L,P) \to (\hat{I},J,\hat{L},P) \to (I',J',L',P') \to Z^n$. So we have $I(S^k;Z^n|I',J') \leq k I(S;\B|\A) + \epsilon$.

Following similar steps and considering $U^k$ instead of $S^k$ and $I',L'$ instead of just $I'$, we can also show that $I(Z^n;U^k|I',L',P') \leq k I(U;\Sp|\Sc,\B) + \epsilon$.
}
\end{appendices}
\end{document}